\documentclass[twocolumn]{aastex63}
\usepackage{chngcntr}

\usepackage{color}
\usepackage{gensymb}
\usepackage{enumitem} 

\newcommand\swift{{\it Swift}}

\newcommand\xmm{{\it XMM-Newton}}

\newcommand\astrosat{{\it AstroSat}}
\newcommand\nustar{{\it NuSTAR}}

\newcommand\s{{\rm~s}}

\newcommand\mpc{{\rm~Mpc}}

\newcommand\kev{{\rm~keV}}
\newcommand\ev{{\rm~eV}}
\newcommand\ergs{{\rm~ergs}}
\newcommand\cm{{\rm~cm}}

\newcommand\angstrom{{\rm~\AA}}

\shortauthors{Tripathi \& Dewangan}

\begin{document}

\title{Thermal Comptonization in a changing corona in the changing-look active galaxy NGC~1566}
\correspondingauthor{Prakash Tripathi}
\email{prakasht@iucaa.in}


\author[0000-0003-4659-7984]{Prakash Tripathi}
\affiliation{Inter University Centre for Astronomy and Astrophysics, Pune, India, 411007}

\author[0000-0003-1589-2075]{Gulab Chand Dewangan}
\affiliation{Inter University Centre for Astronomy and Astrophysics, Pune, India, 411007}

\begin{abstract}
We present  broadband UV/X-ray spectral variability of the changing-look active galactic nucleus NGC~1566 based on  simultaneous near-ultraviolet (NUV) and X-ray observations performed by \xmm{}, \swift{}, and \nustar{} satellites at five different epochs during the declining phase of the 2018 outburst. We found that the accretion disk, soft X-ray excess, and the X-ray power-law components were extremely variable. Additionally, the X-ray power-law flux was correlated with both the soft excess plus disk, and the pure disk fluxes. Our finding shows that at high flux levels the soft X-ray excess and the disk emission both provided the seed photons for thermal Comptonization in the hot corona, whereas at low flux levels where the soft excess was absent, the pure disk emission alone provided the seed photons. The X-ray power-law photon-index was only weakly variable ($\Delta{\Gamma_{hot}}\leq0.06$) and it was not well correlated with the X-ray flux over the declining timescale. On the other hand, we found that the electron temperature of the corona increased from $\sim22$ to $\sim200\kev{}$ with decreasing number of seed photons from June 2018 to August 2019. At the same time, the optical depth of the corona decreased from $\tau_{hot}\sim4$ to $\sim0.7$, and the scattering fraction increased from $\sim1\%$ to $\sim10\%$. These changes suggest structural changes in the hot corona such that it grew in size and became hotter with decreasing accretion rate during the declining phase. The AGN is most likely evolving with decreasing accretion rate towards a state similar to the low/hard state of black hole X-ray binaries. 

\end{abstract}


\section{Introduction}

The primary emission from 
type~1 active galactic nuclei (AGNs) comprises a big-blue-bump in the optical/ultraviolet (UV) band,  soft X-ray excess below $\sim 2\kev{}$, and power-law X-ray continuum with a high energy cut-off { in the range of $\sim 20-1000\kev{}$ \citep[see][]{2015MNRAS.451.4375F, 2018A&A...614A..37T}}. These components are thought to arise from the central engine of AGNs that consists of an accretion disk surrounding the supermassive black-hole (SMBH), and a hot corona. It is believed that the UV photons produced at the innermost regions of the disk via the accretion of matter interact with the hot corona, and are repeatedly up-scattered by the coronal energetic electrons to the X-rays \citep{1980A&A....86..121S, 1991ApJ...380L..51H, 1993ApJ...413..507H}.  
The spectral shape of the Comptonized X-ray emission is well approximated as a power-law with a high energy cut-off. The photon-index of the power-law component is determined by the physical properties such as optical depth and temperature of the corona, whereas the high energy cut-off is generally 2--3 times of the electron temperature \citep[see][]{2000ApJ...540..131P, 2001ApJ...556..716P}. The high energy cut-offs observed in a number of AGNs support the Comptonization scenario in the hot corona \citep[see][]{2015MNRAS.451.4375F,2016MNRAS.458.2454L,2017ApJS..233...17R,2018A&A...614A..37T}. Also, a correlation between the photon-index and the X-ray power-law flux found in several Seyfert type AGNs is interpreted in terms of thermal Comptonization, in which increased seed photons from the disk cool the hot corona \citep[see][]{2003MNRAS.342..355Z, 2009MNRAS.399.1597S}. 

The primary X-ray power-law component then illuminates the disk, interacts with the disk matter, thus gives rise to the iron K features in the 6--7\kev{} band, Compton reflection hump above 10\kev{} \citep{2000PASP..112.1145F},  and perhaps some fraction of the soft X-ray excess below 2\kev{} \citep{2006MNRAS.365.1067C, 2011MNRAS.410.1251N, 2012MNRAS.423.3299N, 2013MNRAS.428.2901W}. Alternatively, the soft X-ray excess may also arise due to  thermal Comptonization of the disk photons in a warm and optically thick corona \citep[see][]{2007ApJ...671.1284D,2012MNRAS.420.1848D, 2015A&A...575A..22M,2018A&A...611A..59P,2020A&A...634A..85P}. 

There are only a few direct observational evidences for thermal Comptonization of the disk seed photons in the hot corona. For example, using one month long IUE and RXTE data on NGC~7469, \citealt{2000ApJ...544..734N} found a correlation between the X-ray power-law photon-index and the UV flux, and interpreted in terms of cooling of the corona with the increasing UV flux \citep[also see][]{2004A&A...413..477P}. \citealt{2013MNRAS.433.1709G} found that the photon-index of the X-ray power-law component increases with the $0.1-1\kev{}$ flux in an X-ray bright radio-loud narrow-line Seyfert~1 (NLS1) galaxy PKS~0558-504, and suggested that the soft X-ray excess is the source of seed photons for the thermal Comptonization. Recently, using simultaneous near-ultraviolet (NUV) and X-ray data of a bright Seyfert~1.2 AGN IC~4329A acquired by \astrosat{},  
\citealt{Tripathi_2021} found the steepening of the X-ray power-law ($\Gamma_{hot}\sim 1.77$ to $\sim 2$) with  increasing NUV flux and interpreted this as the cooling of the corona from $\sim 42\kev{}$ to $\sim 32\kev{}$ (at an optical depth $\tau\sim2.3$). Here, we investigate thermal Comptonization  in a changing-look AGN (CL-AGN) NGC~1566 using multiple sets of broadband UV/X-ray observations. { There is no evidence for variable obscuration in NGC~1566, and its changing-look behavior has been attributed to intrinsic variability \citep[see][]{2019MNRAS.483L..88P,2019MNRAS.483..558O,Tripathi_&_Dewangan}}

 NGC~1566 is the nearest (z = 0.00502) CL-AGN that exhibited an outburst during 2017--2018 with the outburst peak during June--July 2018 (see, \citealt{2019MNRAS.483L..88P,2019MNRAS.483..558O} and references therein). During the outburst, the accretion disk, the soft X-ray excess, and the X-ray power-law fluxes increased by factors of $\sim 30$, $> 200$, and $\sim 25$, respectively, 
 and the accretion disk and the soft excess fluxes have been found to be correlated with the X-ray power-law flux \citep[see][]{Tripathi_&_Dewangan}.  Using the multi-epoch X-ray (0.5--70\kev{}) data of NGC~1566, 
 \citealt{2021MNRAS.507..687J} found that the temperature of the hot corona to increase from $\sim 60\kev{}$ to $\sim 100\kev{}$ during the declining phase of the outburst from June 2018 to August 2019.  
 Here, we study broadband UV--X-ray spectral variability with improved thermal Comptonisation model and accounting for both blurred and distant reflection emission, and then investigate the effect of drastically varying accretion disk and soft X-ray excess emission on the coronal properties during the declining phase of the 2018 outburst. The joint UV/X-ray spectral analyses will reveal not only the nature of the seed photons and their connection to the hot corona, but also the intrinsic variability of the hot corona during the outburst. In particular, the extreme X-ray variability of NGC~1566 will allow us to decouple the optical depth and temperature of the hot corona using the broadband data, and investigate their variations possibly caused by the changes in the corona.

 We organize the paper as follows. We describe the observation and data reduction in section~\ref{sec_obs} followed by broadband UV/X-ray spectral analyses in section~\ref{sec_spectral}. We discuss our results in section~\ref{sec_discuss}, and finally conclude in section~\ref{sec_conclude}.

\section{Observation and data reduction}
\label{sec_obs}

We analyze the simultaneous NUV/X-ray data on NGC~1566 acquired by \xmm{} \citep{2001A&A...365L...1J}, \nustar{} \citep{2013ApJ...770..103H},  and the Neil Gehrels \swift{} Observatory \citep{2004ApJ...611.1005G}.  \citealt{2021MNRAS.507..687J} have analyzed these X-ray data but they did not use the simultaneous NUV data. \citealt{2019MNRAS.483L..88P} and \citealt{Tripathi_&_Dewangan} have  analyzed the \xmm{} data acquired on 26 June 2018 (outburst peak). Here, we use five broadband ($\sim 2000$\angstrom{}  to $80\kev$) data sets acquired with  simultaneous observations at the NUV,  soft and hard X-ray bands. 
We list the five sets of simultaneous UV/X-ray observations in Table~\ref{tab_xrayobs}.

\begin{table*}
\label{tab_xrayobs}
    \centering
    \caption{X-ray observations of NGC~1566}
\begin{tabular}{ccccccccccccc}
    \hline\hline
 Observation&   Date  &\multicolumn{3}{c}{\xmm{}/EPIC-pn} &\multicolumn{3}{c}{\nustar{}/FPMA (B)} &\multicolumn{3}{c}{\swift{}/XRT} \\
    &yyyy/mm/dd    &  ObsID           &T$_{\rm{exp}}^a$   &Rate$^b$            &  ObsID     &T$_{\rm{exp}}^a$ &Rate$^c$                 &  ObsID     &T$_{\rm{exp}}^a$ &Rate$^d$ \\
         
       \hline
 Obs1      &2018/06/26     &0800840201 &65   &$10.24$  &80301601002    & 57  &1.59 (1.54)   &               &   &  \\
 Obs2      &2018/10/04     &0820530401 &74   &$5.88$   &80401601002    & 75  &0.45 (0.42)   &               &   &    \\
 Obs3      &2019/06/05     &0840800401 &64   &4.41    &80502606002    & 57  &0.29 (0.28)  &               &   &    \\
 Obs4      &2019/08/18     &           &   &   &60501031004    &77  & 0.27 (0.26)  &00088910002    &1.6  &$0.25$ \\
 Obs5      &2019/08/21     &           &   &   &60501031006    &86   &0.29 (0.28)   &00088910003    &1.9   &$0.18$   \\
\hline\hline
    \end{tabular}
    \\
    $^a$ The net exposure time in ks\\
    $^b$ The net source count rate in the unit of count\s$^{-1}$ in the 0.5--10\kev{} band \\
    $^c$ The net source count rate in the unit of count\s$^{-1}$ in the 3--78\kev{} band \\
    $^d$ The net source count rate in the unit of count\s$^{-1}$ in the 0.5--7\kev{} band \\
   
\end{table*}

\subsection{\xmm{} data}

We reduced the \xmm{} 
data using the Science Analysis System (SAS~v18.0.0) software and the latest calibration files. { The pn European Photon Imaging Camera (EPIC-pn; \citealt{2001A&A...365L..18S}) was operated in the small window mode with thick filter during the June and October 2018 observations, and in the small window mode with medium filter during the June 2019 observation.} We reprocessed the 
EPIC-pn data with {\textsc{epproc}} task. We did not notice strong flaring particle background in any of the EPIC-pn observations, therefore we did not filter the data for particle background. { Further, we checked for the pile-up using {\textsc{epatplot}} task, and noticed pile-up in the June 2018 data only.} Following \citealt{2021MNRAS.507..687J}, we 
corrected for the pile-up by excluding events in an inner circular region of $10\arcsec$ radius from the clean event lists. 
We extracted source spectrum from a circular region of $30\arcsec$ radius centred at the source position and background spectrum from a circular region of $40\arcsec$ radius from a source-free region for each observation. We generated the redistribution matrix (RMF) and ancillary response files (ARF) using {\textsc{rmfgen}} and {\textsc{arfgen}} tasks, respectively. We grouped the spectra to a minimum of 25 counts/bin and {\textsc{oversample = 5}} using {\textsc{specgroup}} task. We list the net source count rate and the exposure time in Table~\ref{tab_xrayobs} for each observation.

We also processed the Optical Monitor (OM; \citealt{2001A&A...365L..36M}) data from the three observations using the {\textsc{omichain}} task. 
We performed aperture photometry on the clean UVW2 ($\rm{\lambda_{eff}} =2120{\rm{\AA}}$, $\rm{\Delta{\lambda}} = 500{\rm{\AA}}$) and UVM2 ($\rm{\lambda_{eff}} =2310 {\rm{\AA}}$, $\rm{\Delta{\lambda}} = 480 {\rm{\AA}}$) images of the source, and derived the net source count rate using the {\textsc{omsource}} task. We then corrected the observed net source count rate for the Galactic extinction and the BLR/NLR contributions, and derived the intrinsic count rates of the source following the same procedure as mentioned in \citealt{Tripathi_&_Dewangan}. We wrote these intrinsic count rates in the OGIP compliant spectral files derived using {\textsc{om2pha}} task.

\subsection{\nustar{} data}
We processed the \nustar{} 
data with  {\textsc{nustardas} version 2.1.1} and the latest calibration database (CALDB, version 20210728). We generated the clean event files for each observation using {\textsc{nupipeline}} task. We extracted the source spectrum from a circular region of $60\arcsec$ radius centered at the source position and background spectrum from a source free circular region of $90\arcsec$ radius using {\textsc{nuproduct}} task. 
We grouped each spectrum to have at least 25 counts per spectral bin. The net source count rate and the exposure time are listed in Table~\ref{tab_xrayobs} for each observation.

\subsection{\swift{} data}

We processed the X-Ray Telescope (XRT; \citealt{2005SSRv..120..165B}) data using the standard online tool (user\_objects\footnote{\url{https://www.swift.ac.uk/user_objects/}}) developed by the \swift{}t Science Data Centre, UK \citep{2009MNRAS.397.1177E}, and obtained the spectral data in  
the Photon-Counting (PC) mode. 
We grouped each spectral data set to have at least 20 counts per bin using {\textsc{grppha}} task. The net source count rate and the exposure time are listed in Table~\ref{tab_xrayobs}. 

Further, we also used the clean NUV (below 3000\AA) images of the source observed by the Ultra-Violet Optical Telescope (UVOT; \citealt{2005SSRv..120...95R}).
We used the {\textsc{uvotsource}} task and  performed the aperture photometry  on the clean,  processed images  of the source in the UVW2 ($\rm{\lambda_{eff}} =1928 {\rm{\AA}}$, $\rm{\Delta{\lambda}} = 657 {\rm{\AA}}$) and UVM2 ($\rm{\lambda_{eff}} =2246 {\rm{\AA}}$, $\rm{\Delta{\lambda}} = 498 {\rm{\AA}}$) bands. 
We extracted source counts from a circular region of $10\arcsec$ radius centered at the source position, and background counts from a similar size of nearby annular region. We corrected the UVOT count rates for the Galactic extinction. 
We estimated the fractional BLR/NLR contribution for the UVW2 and UVM2 filters to be $\sim 23\%$ using the effective areas of these filters (see \citealt{Tripathi_&_Dewangan}). We subtracted these contributions from the Galactic extinction corrected count rates, and derived the intrinsic source count rates. We wrote these intrinsic count rates in the OGIP compliant spectral files generated by the {\textsc{uvot2pha}} task. 

\section{Spectral analyses}
\label{sec_spectral}

We performed spectral analyses using {\textsc{xspec}}~(v12.12.0) \citep{1996ASPC..101...17A}. We used the $\chi^2$-minimization technique and quoted  $1\sigma$ errors on the best-fit spectral parameters. For the Galactic absorption, we used the {\textsc{tbabs}} model in {\textsc{xspec}} with a fixed equivalent hydrogen column density of $N_H = 9.19\times10^{19}\cm^{-2}$ \citep{2005A&A...440..775K}. We used the abundance and absorption cross-sections from \cite{2009ARA&A..47..481A} and \cite{1996ApJ...465..487V}, respectively.

\begin{table*}
	\centering
	\caption{The best-fit parameters with $1\sigma$ errors of NGC~1566 derived from the NUV/X-ray SEDs using {\textsc{constant$\times$tbabs$\times$zxipcf$\times$(optxagnf+ relxill+xillver)}} model. The parameters with (f) sign are fixed.}
	\label{tab_nuvxrayspec_1}
	\begin{tabular}{ccccccc}
		\hline\hline
		Model& Parameter       &Obs1  &Obs2   &Obs3 &Obs4  &Obs5\\
		\hline
		{\textsc{const}} & $C_{\rm{PN}}$   &1 (f)   & 1 (f) &1(f)   & -- &--\\
		& $C_{\rm{FPMA}}$   &$1.02^{+0.01}_{-0.01}$ &$1.07^{+0.01}_{-0.01}$ &$1.04^{+0.01}_{-0.01}$  &$1.0^{+0.06}_{-0.05}$ &$1$ (f)\\
		 & $C_{\rm{FPMB}}$   &$1.06^{+0.01}_{-0.01}$ &$1.09^{+0.01}_{-0.01}$ &$1.1^{+0.01}_{-0.01}$  &$1.04^{+0.06}_{-0.05}$ &$1.05^{+0.01}_{-0.01}$\\
		 & $C_{\rm{XRT}}$   &--&-- &--& 1 (f) &$0.61^{+0.04}_{-0.03}$\\
		\\\
		{\textsc{zxipcf}} & $N_{{H}} (10^{22}\cm^{-2})$   &$3.0^{+0.6}_{-0.3}$&--&--&--&--\\
		                  & $\log({\xi})$   &$1.45^{+0.2}_{-0.2}$&--&--&--&--\\
	                	& $CF$   &$0.16^{+0.06}_{-0.02}$&--&--&--&--\\
		\\\
		{\textsc{optxagnf}} &$\log(L/L_{Edd})$ &$-1.44^{+0.01}_{-0.01}$ &$-2.11^{+0.01}_{-0.01}$    &$-2.38^{+0.01}_{-0.01}$  &$-2.42^{+0.02}_{-0.02}$    &$-2.407^{+0.005}_{-0.005}$\\
		                    &$R_{cor} (GM_{BH}/c^2)$ & $72^{+3}_{-2}$ & $>88$  &$>75$ &$59^{+3}_{-3}$ & $69^{+2}_{-1}$\\
	                    	&$kT_{w}$ (keV) &$0.60^{+0.03}_{-0.04}$ &$0.41^{+0.04}_{-0.03}$ &$0.33^{+0.09}_{-0.09}$ &--&--\\
	                    	&$\tau_{w}$   &$9.3^{+0.6}_{-0.4}$    &$11.5^{+0.5}_{-0.5}$ &$13^{+4}_{-2}$ &--&--\\
	                    	&$\Gamma_{hot}$ & $1.43^{+0.01}_{-0.01}$ &$1.68^{+0.01}_{-0.01}$ &$1.71^{+0.01}_{-0.01}$   &$1.72^{+0.01}_{-0.01}$ &$1.73^{+0.01}_{-0.01}$\\
	                    	&$f_{PL}$ &$\leq 0.05$ &$0.69^{+0.02}_{-0.01}$ &$0.838^{+0.092}_{-0.005}$ &1 (f)   & 1 (f)\\
		                 \\\
	 {\textsc{relxill}}    
	                        &$\log({\xi_{rel}})$   &$4.1^{+0.02}_{-0.01}$ & --   &--&--&--\\
	                        &$N_{rel} (10^{-5})$  &$11.1^{+0.2}_{-0.7}$  &-- &--& --   &--\\

	 \\\
		 {\textsc{xillver}} 
		                    &$A_{Fe}$    &4.6 (f)  &$4.6^{+2.1}_{-0.7}$ &4.6 (f)    &4.6 (f)    &4.6 (f)\\
	                        &$E_{cut}$ (keV)   &300 (f) &  300 (f)  &300 (f)   &300 (f)   &300 (f)  \\
		 &$N_{xill} (10^{-5})$  &$8.0^{+0.6}_{-0.6}$  &$3.9^{+0.4}_{-0.5}$ &$1.9^{+0.1}_{-0.1}$ &$2.5^{+0.3}_{-0.3}$ &$2.0^{+0.2}_{-0.2}$ \\
 \hline
&	$\chi^2/dof$						&1676/1547  &1258/1189  &981/931    &700/763    &790/813 \\
		\hline\hline									
	\end{tabular}
\end{table*}

We began our spectral analyses by fitting the OM, EPIC-pn (0.5--10\kev{}), and the \nustar{}/FPMA \& FPMB (3--70\kev{}) spectral data from the 26 June 2018 epoch (Obs1). We did not use the \nustar{} data above 70\kev{} due to the poor signal-to-noise ratio. { We used a broadband continuum model {\textsc{optxagnf}} \citep{2012MNRAS.420.1848D} to fit the disk, soft X-ray excess, and the X-ray power-law components of the broadband UV/X-ray spectra. The main parameters of this model are mass of the SMBH ($M_{BH}$ in the unit of the solar mass $M_{\odot}$), comoving distance of the source ($D$ in\mpc{}), Eddington ratio ($\log{\dot{m}} = \log{L/L_{Edd}}$), dimensionless spin parameter of the SMBH ($a$), radius of the warm corona ($R_{cor}$) in units of gravitational radius $R_g = GM_{BH}/c^2$, outer disk radius ($R_{out}$ in $R_g$), temperature ($kT_{w}$ in\kev{}) and optical depth ($\tau_{w}$) of the warm corona, photon-index of the X-ray power-law component ($\Gamma_{hot}$), and the fraction of the power below $R_{cor}$ emitted as the hard Comptonization component ($f_{PL}$). The {\textsc{optxagnf}} model utilizes the {\textsc{nthcomp}} \citep{nthcomp1} model to fit the X-ray power-law component assuming the electron temperature of the hot corona fixed at $kT_{e} = 100\kev{}$ \citep[see][]{2012MNRAS.420.1848D}. We fixed the black-hole mass $M_{BH}=8.32\times10^{6}M_{\odot}$ \citep{2002ApJ...579..530W}, the comoving distance $D=21.3\mpc{}$ \citep{2019MNRAS.487.2797E}, and outer disk radius $R_{out}=10^3R_g$. We also fixed the black-hole spin at the reported value of $a=0$ by \cite{2019MNRAS.483L..88P} and \cite{2021MNRAS.507..687J}. 

Next, we used the {\textsc{relxill}} \citep{2014ApJ...782...76G} and the {\textsc{xillver}} \citep{2013ApJ...768..146G} model to fit the blurred and distant reflection features, respectively, as reported by \cite{2019MNRAS.483L..88P}, \cite{2021MNRAS.507..687J}, and  \cite{Tripathi_&_Dewangan}. The main parameters of the {\textsc{relxill}} model are emissivity index ($q$ assuming a single emissivity profile $\epsilon\propto R^{-q}$, where $R$ is the radial distance), the black-hole spin ($a$), inclination angle of the accretion disk ($\theta$), inner and outer disk radii ($R_{in}$ and $R_{out}$ in $R_g$), photon-index ($\Gamma_{hot}$) of the incident X-ray power-law component, ionization parameter ($\log({\xi_{rel}/\ergs{}\cm{}\s^{-1}})$), iron abundance ($A_{Fe}$ in the solar unit), high energy cut-off ($E_{cut}$ in\kev{}), reflection fraction ($RF_{rel}$), and normalization ($N_{rel}$). The blurred reflection component ({\textsc{relxill}}) is weak in this source (see \citealt{2019MNRAS.483L..88P} and \citealt{Tripathi_&_Dewangan}), we therefore fixed the emissivity index at $q=3$, the black-hole spin at $a=0$, inclination angle at $\theta=10^{\degree}$, and the inner disk radius at $R_{in}=6R_g$ as reported by \cite{2019MNRAS.483L..88P}. We also fixed the outer disk radius at $R_{out}=10^3 R_g$, ionization parameter of the {\textsc{xillver}} model at $\log{\xi_{xill}}=0$, reflection fraction of the {\textsc{relxill}} and {\textsc{xillver}} models at $RF_{rel}=-1$ and $RF_{xill}=-1$ to fit the reflection spectrum only, and the iron abundance of both the reflection models at the best-fit value derived from the 04 October 2018 epoch ($A_{Fe}=4.6$, see below and Table~\ref{tab_nuvxrayspec_1}). We tied the photon index of the reflection models with that of the {\textsc{optxagnf}} model, and fixed the cut-off energy at $E_{cut} = 300 keV$ assuming $E_{cut}=3kT_e$ where $kT_e$ is fixed at $100\kev{}$ for the {\textsc{optxagnf}} model \citep{2012MNRAS.420.1848D}. We allowed the $\log{\xi_{rel}}$ and the normalizations of the reflection models $N_{rel}$, and $N_{xill}$ to vary freely. We also fitted the weak warm absorption features present in this epoch \citep[see][]{2019MNRAS.483L..88P} using a {\textsc{zxipcf}} model. We used a {\textsc{constant}} model for the cross-normalization between the EPIC-pn and the \nustar{} data. We fixed this parameter at $1$ for the EPIC-pn and the OM data, and allowed it to vary for the \nustar{} (FPMA \& FPMB) data. Thus, our model is {\textsc{constant$\times$tbabs$\times$zxipcf$\times$(optxagnf+ relxill+xillver)}} in {\textsc{xspec}} terminology. The fit with this model resulted in an statistically acceptable fit with $\chi^2 = 1675$ for 1547 degrees of freedom ($dof$). 

We also fitted the OM, EPIC-pn (0.5--10\kev{}), and \nustar{}/FPMA \& FPMB (3--70\kev{}) data from the 04 October 2018 epoch (Obs2). We used the model
{\textsc{constant$\times$tbabs$\times$(optxagnf+xillver)}}. These data did not require the blurred reflection and the warm absorption components. We allowed to vary the parameters of the {\textsc{optxagnf}} model as before, and the iron abundance ($A_{Fe}$) and normalization ($N_{xill}$) of the {\textsc{xillver}} model. The fit resulted in $\chi^2/dof = 1258/1189$. 

Similarly, we fitted the OM, UVOT, EPIC-pn (0.5--10\kev{}), and \nustar{} (3--50\kev{}) data from the 05 June 2019 epoch (Obs3). The source was at a low flux state in 2019 epochs, and the SNR of the \nustar{} data above 50\kev{} was poor. We therefore used the \nustar{} data below 50\kev{} for all the 2019 epochs. 
We used the model {\textsc{constant$\times$tbabs$\times$(optxagnf+xillver)}} to fit the broadband NUV/X-ray data. As before, we allowed to vary the parameters of the {\textsc{optxagnf}} model, and the normalization of the {\textsc{xillver}} component. The fit resulted in $\chi^2/dof=981/931$.

Further, we fitted the UVOT, XRT (0.5--7\kev{}), and \nustar{} (3--50\kev{}) data from the 18 August 2019 epoch (Obs4) using the {\textsc{constant$\times$tbabs$\times$(optxagnf+xillver)}} model. We did not notice the soft X-ray excess component in this epoch, and therefore fixed the $f_{PL}$ parameter of {\textsc{optxagnf}} model at 1. 
We allowed other parameters to vary freely, as before. Here, we fixed the cross-normalization parameter at $1$ for the XRT and UVOT data, and allowed it to vary for the \nustar{} data. The fit resulted in $\chi^2/dof=700/763$.

Finally, we fitted the UVOT, XRT (0.5--6\kev{}), and \nustar{} (3--50\kev{}) data of the 21 August 2019 epoch (Obs5) with {\textsc{constant$\times$tbabs$\times$(optxagnf+xillver)}} model. Here also, we fixed the $f_{PL}$ at 1 as the soft X-ray excess was absent. 
While fitting, we found that the cross-normalization parameter ({\textsc{constant}}) for the \nustar{} data to be $\sim 1.4$ relative to the XRT data. This is because the \swift{} observations were performed at the beginning of the \nustar{} observations when the source flux was $\sim 1.4$ times lower than the maximum value (see the XRT and \nustar{}/FPMA lightcurves in Figure~\ref{fig_lc}).  
Therefore, to calculate the best-fit parameters 
based on the long exposure \nustar{} data of the 21 August 2019 epoch, we fixed the {\textsc{constant}} at 1 for the \nustar{}/FPMA data, and allowed it to vary for the \nustar{}/FPMB and XRT data. 
The fit resulted in $\chi^2/dof=790/813$. The best-fit parameters for the five-epochs are listed in Table~\ref{tab_nuvxrayspec_1}. As it can be seen that the normalization of the {\textsc{xillver}} model is not constant across the five epochs, especially for the first epoch this parameter is $\sim 4$ times larger than those of the 2019 epochs. Also, the photon-index of the X-ray power-law is low ($\Gamma_{hot}\sim 1.43$) for the first epoch. One reason could be the fixed electron temperature of the hot corona in the {\textsc{optxagnf}} model. We therefore further analysed the spectral data using more physically motivated models. 
}

\begin{figure*}
    \centering
   \includegraphics[scale=0.35]{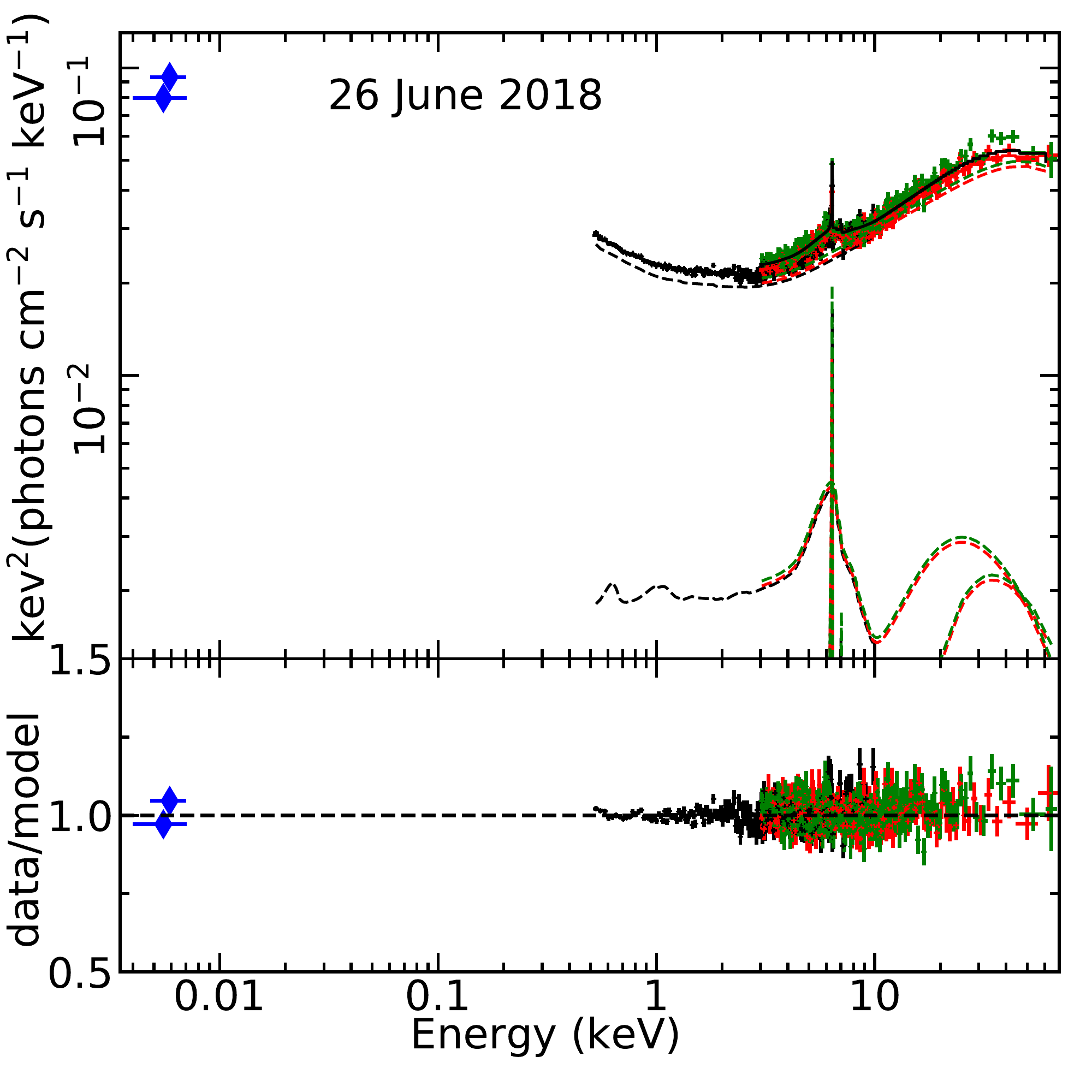}
   \includegraphics[scale=0.35]{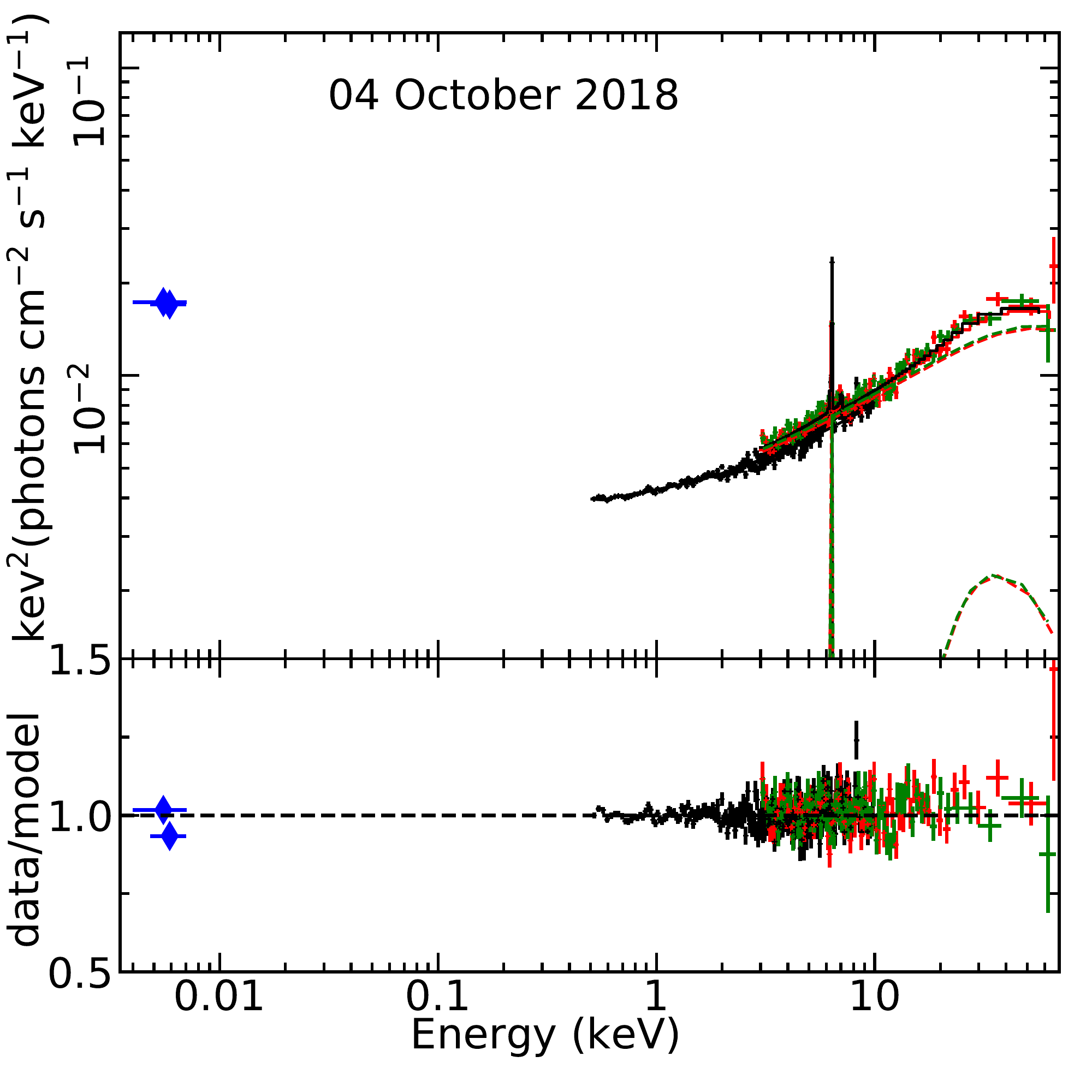}
   \includegraphics[scale=0.35]{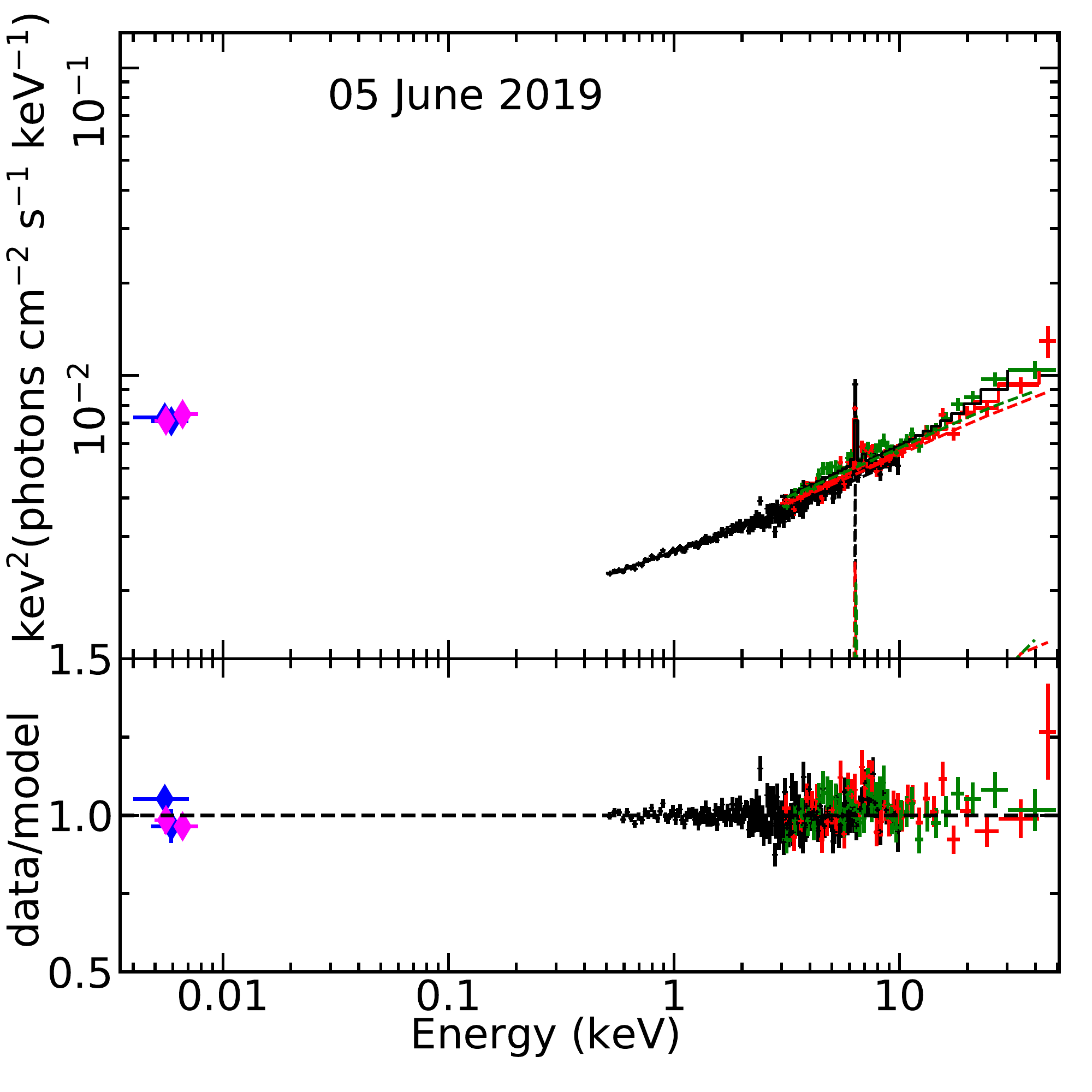}
   \includegraphics[scale=0.35]{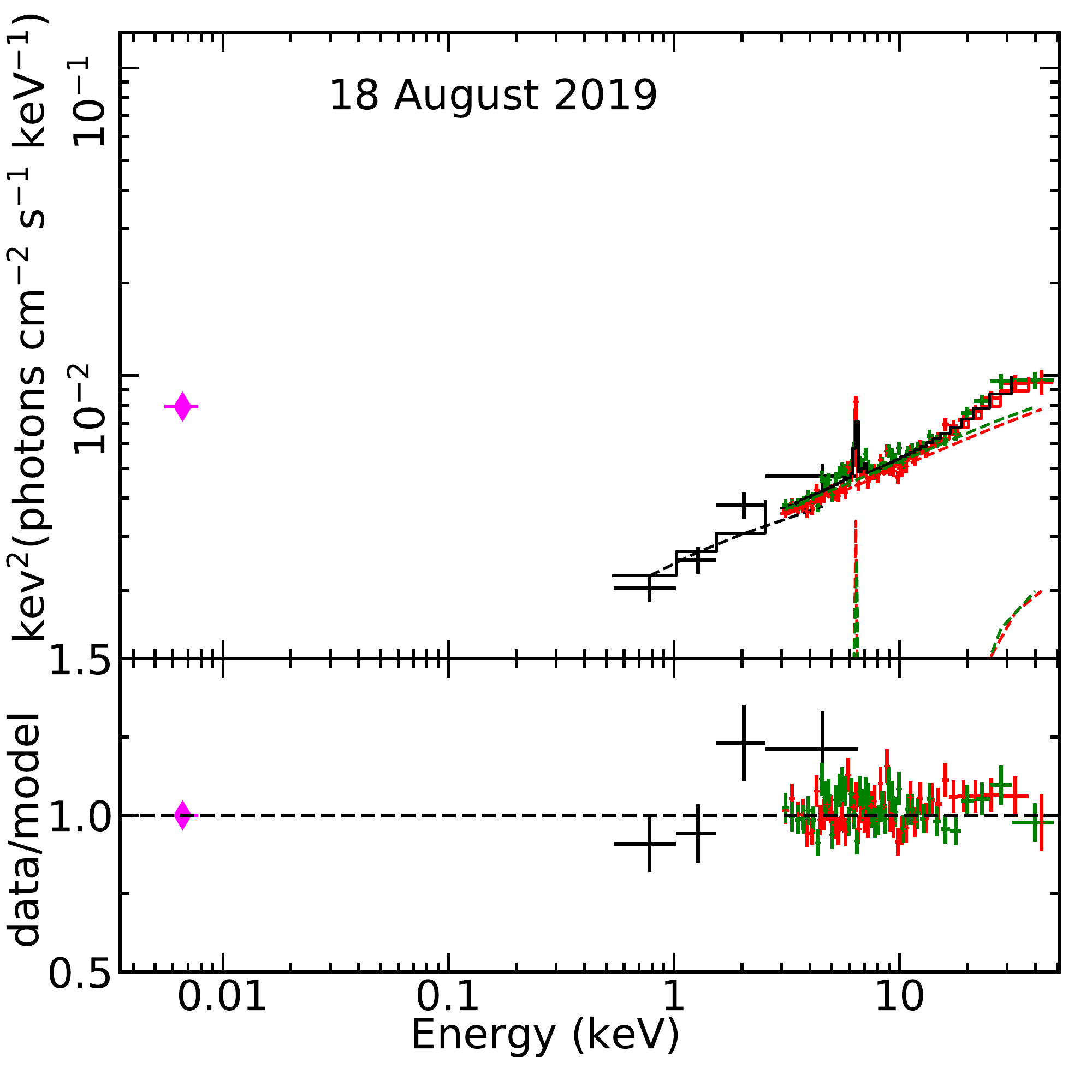}
   \includegraphics[scale=0.35]{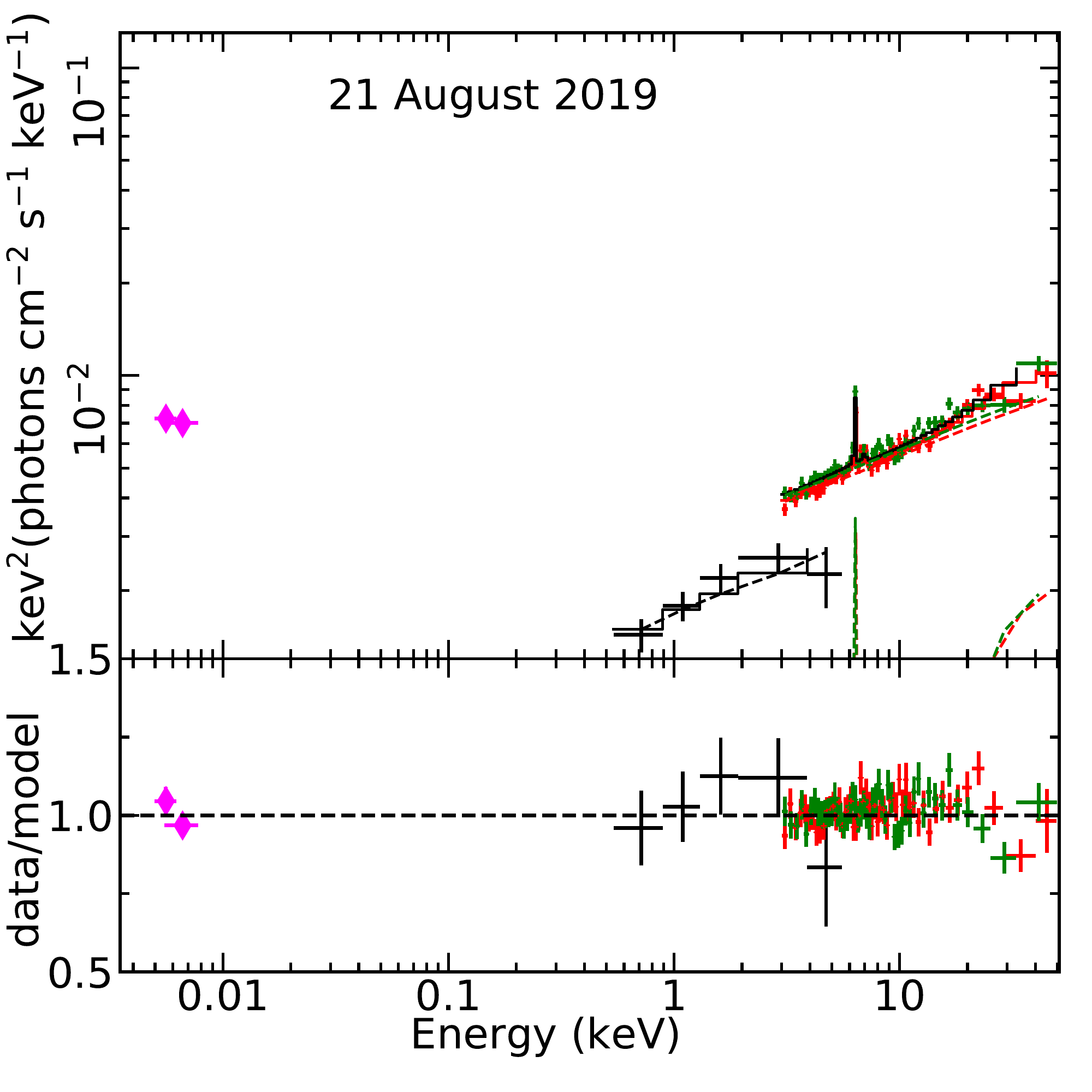}
    \caption{The unfolded NUV and X-ray spectral data, the best-fitting models, and the data-to-model ratios for the five simultaneous observations. 
    The EPIC-pn and XRT (black plus), FPMA (red plus), FPMB (green plus), OM (blue diamond), and UVOT (magenta diamond) data are shown. The best-fitting model components are shown in the dashed lines. 
    For clarity purpose, the data have been rebinned.}
    \label{fig_sed}
\end{figure*}

Here, our main motivation is to investigate variations in the hot corona i.e., variations in the coronal temperature and optical depth and any connection with the accretion rate and seed photon flux  during the declining phase of the 2018 outburst when all emission components varied strongly \citep{Tripathi_&_Dewangan}. 
We therefore used the thermal Comptonization model {\textsc{thcomp}} \citep{2020MNRAS.492.5234Z} to describe the broadband X-ray power-law continuum. { As before, we first fitted the 26 June 2018 data.} We assumed that both the thermal emission from the accretion disk and the soft X-ray excess components to provide the  seed photons for thermal Comptonization in the hot corona. We therefore used the {\textsc{optxagnf}} \citep{2012MNRAS.420.1848D} model to fit the accretion disk and the soft X-ray excess components only by fixing the $f_{PL}=0$. 
We allowed the Eddington ratio, temperature and optical depth of the warm corona, and the warm corona radius ($R_{cor}$) to vary freely. We then Comptonized the disk and the soft excess emission by convolving with the thermal Comptonization  model {\textsc{thcomp}} \citep{2020MNRAS.492.5234Z}, thus allowing us to describe the X-ray power-law component and determine the coronal properties. To use the {\textsc{thcomp}} convolution model,  we extended the energy grid from $10^{-5}\kev{}$ to $500\kev{}$ with the {\textsc{xspec}} command {\textsc{energies}}. We allowed to vary the \textsc{thcomp} parameters  $\Gamma_{hot}$ (photon index of the Comptonized X-ray power-law),  $kT_e$ (electron temperature of the hot corona), and  $f_{sc}$  (the scattering fraction where $0\leq f_{sc}\leq1$ describing the fraction of the seed photons that are up-scattered). We chose $\Gamma_{hot}$ instead of the optical depth $\tau_{hot}$ of the corona to vary. { We also used the {\textsc{relxill}} and the {\textsc{xillver}} models to fit the blurred and the distant reflection features, and the {\textsc{zxipcf}} model to fit the warm absorption. Thus, our model is {\textsc{constant$\times$tbabs$\times$zxipcf$\times$(thcomp$*$optxagnf+ relxill+xillver)}}. The fit with this model resulted in $\chi^2/dof = 1649/1546$. This fit is slightly better than the previous one when the cut-off energy was not a free parameter.} 

\begin{table*}
	\centering
	\caption{The best-fit parameters with $1\sigma$ errors of NGC~1566 derived from the NUV/X-ray SEDs using {\textsc{constant$\times$tbabs$\times$zxipcf$\times$(thcomp$*$optxagnf+ relxill+xillver)}} model. Here, errors have been calculated using the MCMC chain. The parameters with (f) sign are fixed.}
	\label{tab_nuvxrayspec}
	\begin{tabular}{ccccccc}
		\hline\hline
		Model& Parameter       &Obs1  &Obs2   &Obs3 &Obs4  &Obs5\\
		\hline
		{\textsc{const}} & $C_{\rm{PN}}$   &1 (f)   & 1 (f) &1(f)   & -- &--\\
		& $C_{\rm{FPMA}}$   &$1.02^{+0.01}_{-0.01}$ &$1.07^{+0.01}_{-0.01}$ &$1.04^{+0.01}_{-0.01}$  &$1.0^{+0.07}_{-0.05}$ &$1$ (f)\\
		 & $C_{\rm{FPMB}}$   &$1.06^{+0.01}_{-0.01}$ &$1.09^{+0.01}_{-0.01}$ &$1.09^{+0.01}_{-0.01}$  &$1.04^{+0.07}_{-0.05}$ &$1.05^{+0.01}_{-0.01}$\\
		 & $C_{\rm{XRT}}$   &--&-- &--& 1 (f) &$0.61^{+0.04}_{-0.03}$\\
		\\\
		{\textsc{zxipcf}} & $N_{{H}} (10^{22}\cm^{-2})$   &$1.09^{+0.24}_{-0.05}$&--&--&--&--\\
		                  & $\log({\xi})$   &$-0.4^{+0.5}_{-0.1}$&--&--&--&--\\
	                	& $CF$   &$0.22^{+0.04}_{-0.01}$&--&--&--&--\\
		\\\
{\textsc{thcomp}} & $\Gamma_{hot}$   &$1.66^{+0.01}_{-0.01}$    &$1.67^{+0.01}_{-0.01}$ &$1.70^{+0.01}_{-0.01}$    &$1.71^{+0.01}_{-0.01}$ &$1.72^{+0.01}_{-0.01}$ \\
		                 &$kT_e$ (keV)  &$22.7^{+2.0}_{-1.5}$   &$27^{+9}_{-2}$   &$57^{+11}_{-18}$ &$133^{+101}_{-46}$  &$207^{+73}_{-49}$  \\
		                &$f_{sc} (10^{-2})$   &$0.91^{+0.06}_{-0.04}$   &$1.92^{+0.12}_{-0.07}$   &$5.3^{+0.3}_{-0.3}$    &$7.0^{+0.9}_{-0.9}$    &$9.9^{+0.8}_{-0.9}$ \\
		                &$\tau_{hot}$ &$4.0^{+0.2}_{-0.2}$    &$3.5^{+0.7}_{-0.2}$    &$2.1^{+0.6}_{-0.3}$ &$1.1^{+0.4}_{-0.4}$    &$0.7^{+0.2}_{-0.2}$\\
		                & $f_{xpl} (10^{-10})$    &$2.21^{+0.05}_{-0.05}$ &$0.64^{+0.03}_{-0.03}$    &$0.52^{+0.05}_{-0.05}$    &$0.52^{+0.03}_{-0.03}$    &$0.58^{+0.02}_{-0.02}$\\
		                \\\
		{\textsc{optxagnf}} &$\log(L/L_{Edd})$ &$-1.36^{+0.01}_{-0.004}$ &$-2.285^{+0.005}_{-0.005}$    &$-2.81^{+0.01}_{-0.01}$  &$-2.75^{+0.08}_{-0.06}$    &$-2.79^{+0.05}_{-0.06}$\\
		                    &$R_{cor} (GM_{BH}/c^2)$ & $56^{+2}_{-3}$ &$17^{+2}_{-1}$   & $14^{+2}_{-1}$  &$19^{+10}_{-10}$  &$19^{+7}_{-9}$ \\
	                    	&$kT_{w}$ (keV) &$0.57^{+0.02}_{-0.02}$ &$0.35^{+0.03}_{-0.02}$ &$0.37^{+0.06}_{-0.05}$ &--&--\\
	                    	&$\tau_{w}$   &$9.1^{+0.2}_{-0.2}$    &$11.6^{+0.4}_{-0.6}$ &$12^{+1}_{-1}$ &--&--\\
		                & $f_{seed} (10^{-10})$    &$10.66^{+0.27}_{-0.15}$ &$1.13^{+0.06}_{-0.09}$    &$0.32^{+0.01}_{-0.01}$    &$0.25^{+0.17}_{-0.14}$    &$0.23^{+0.01}_{-0.01}$\\
		                 & $f_{disk} (10^{-10})$    &$3.0^{+0.2}_{-0.1}$ &$0.81^{+0.04}_{-0.06}$    &$0.26^{+0.02}_{-0.02}$    &$0.25^{+0.17}_{-0.14}$&$0.23^{+0.01}_{-0.01}$\\
		           & $L_{Bol}/L_{Edd}$ (\%)    & $6.88$   &$1.05$   &$0.53$   &$0.51$& $0.54$\\
		                 \\\
	 {\textsc{relxill}}    
	                        &$\log({\xi_{rel}})$   &$3.36^{+0.04}_{-0.07}$ & --   &--&--&--\\
	                        &$N_{rel} (10^{-5})$  &$1.5^{+0.2}_{-0.1}$  &-- &--& --   &--\\

	 \\\
		 {\textsc{xillver}} 
		                    &$A_{Fe}$    &5.1 (f)  &$5.1^{+1.4}_{-0.6}$ &5.1 (f)    &5.1 (f)    &5.1 (f)\\
		 &$N_{xill} (10^{-5})$  &$2.8^{+0.4}_{-0.4}$  &$2.6^{+0.3}_{-0.2}$ &$1.6^{+0.2}_{-0.2}$ &$2.5^{+0.6}_{-0.3}$ &$2.5$ (f)\\
 \hline
&	$\chi^2/dof$						&1644/1546  &1256/1188  &982/930    &700/761    &795/812 \\
		\hline\hline									
	\end{tabular}
	\\$f_{seed}$ = the {\textsc{optxagnf}} model flux (disk + soft excess) in the $1\rm{\mu}-2\kev{}$ band in the unit of \ergs{}\cm$^{-2}$\s$^{-1}$\\
	$f_{xpl}$ = the {\textsc{thcomp$*$optxagnf}} model flux in the $2-500\kev{}$ band in the unit of \ergs{}\cm$^{-2}$\s$^{-1}$\\
	$f_{disk}$ = the pure disk flux in the $1\rm{\mu}-2\kev{}$ band in the unit of \ergs{}\cm$^{-2}$\s$^{-1}$ derived using the {\textsc{optxagnf}} parameters\\
	$L_{Bol}/L_{Edd}$ = the Eddington ratio derived from the bolometric luminosity $L_{Bol}$ in the $1\rm{\mu}-500\kev{}$ band.\\
\end{table*}

{Similarly, we fitted the UV/X-ray spectral data from other four epochs following the same procedure mentioned above. Each time, we convolved the {\textsc{thcomp}} model with the {\textsc{optxagnf}} model to fit the X-ray power-law component.} We have shown the best-fit models, spectral data, and the data to model ratios in Fig.~\ref{fig_sed} for all five epoch. We used the Morkov Chain Monte Carlo in {\textsc{xspec}} to calculate the errors on the best-fit parameters. We used the Goodman-Weare algorithm with 200 walkers and a total length of $5\times10^{4}$. We discarded first few thousands steps ($5000-8000$) of the chain to acquire steady state. We calculated $1\sigma$ errors using the chain.  The best-fit parameters with the errors are listed in Table~\ref{tab_nuvxrayspec}. { The normalization of the {\textsc{xillver}} model is not variable across the five epochs, and the value of the X-ray power-law photon-index for the first epoch seems more reasonable (see Table~\ref{tab_nuvxrayspec}).} We have shown the contour plots for one epoch (04 October 2018) in Fig.~\ref{fig_mcmc_contours}.

Further, we calculated the disk plus soft X-ray excess (seed) flux ($f_{seed}$) from the best-fit {\textsc{optxagnf}} model in the $1\rm{\mu}-2\kev{}$ band { (where $1\rm{\mu}\sim 1.24\ev{}$ is the low energy end of the BBB component)}, and the X-ray power-law flux ($f_{xpl}$) from the {\textsc{thcomp$*$optxagnf}} model in the $2-500\kev{}$ band. Using the best-fit parameters ($\log\dot{m}$ and $R_{cor}$) of the {\textsc{optxagnf}} model, we also calculated the pure disk flux ($f_{disk}$) for June 2018, October 2018, and June 2019 epochs where the soft X-ray excess was also present. We also derived the intrinsic bolometric luminosity ($L_{Bol}$) of the source using the best-fit model (after removing the Galactic and the internal absorption components) in the $1\rm{\mu}-500\kev{}$ band, and converted this into the Eddington ratio as $L_{Bol}/L_{Edd}$ for each epoch. We list these fluxes and Eddington ratios in Table~\ref{tab_nuvxrayspec}. 

We derived the optical depth of the corona using the best-fit electron temperature ($kT_e$) and the X-ray power-law photon-index ($\Gamma_{hot}$) (with their $1\sigma$ errors) using the following equation \citep{nthcomp1,nthcomp2}
\begin{equation}
\label{eqn1}
    \tau_{hot} = \sqrt{\frac{9}{4}+\frac{3m_e{c^2}}{kT_e[(\Gamma_{hot}+\frac{1}{2})^2-\frac{9}{4}]}} - \frac{3}{2}
\end{equation}
where, $m_e$ is the mass of electron and $c$ is the speed of light. We have listed the optical depth of the corona in Table~\ref{tab_nuvxrayspec} which decreases from $\sim 4$ to $\sim 0.7$ during the declining phase of the outburst. Independently, we also calculated the optical depth of the hot corona by fitting the broadband spectral data of five epochs using the {\textsc{thcomp}} model. In the negative parameters space ({ for values less than 0}), the $\Gamma_{hot}$ parameter of the {\textsc{thcomp}} model turns into the optical depth. We found the optical depth values derived from the two different methods to be consistent.  

\section{Results and Discussion}
\label{sec_discuss}

We analyzed the simultaneous NUV/X-ray data on  NGC~1566 
acquired by \xmm{}, \swift{}, and \nustar{} from June 2018 to August 2019 at five different epochs. 
We found that the NUV/X-ray spectra of the source consist of a variable X-ray power-law with photon-index $\Gamma_{hot}\sim 1.66-1.72$, soft X-ray excess emission below 2\kev{}, broad/narrow iron lines in the 6--7\kev{} band, a weak Compton reflection hump above 10\kev{}, warm absorbing component, and the accretion disk emission (BBB), thus confirming  previous results \citep[see][]{2019MNRAS.483L..88P, 2019MNRAS.483..558O, 2021MNRAS.507..687J, Tripathi_&_Dewangan}. The photon indices of the X-ray power-law component derived in our analysis for the five epochs are consistent with those derived in \citealt{2021MNRAS.507..687J} by fitting the soft X-ray excess and the X-ray power-law components simultaneously using the {\textsc{optxagnf}} model. As we discuss below, our results on the broadband UV/X-ray spectral variability derived using the improved Comptonization model and accounting for both blurred and distant reflection have allowed us to probe the connection between the accretion disk and changes in the coronal properties.


\begin{figure}
    \centering
   \includegraphics[scale=0.42]{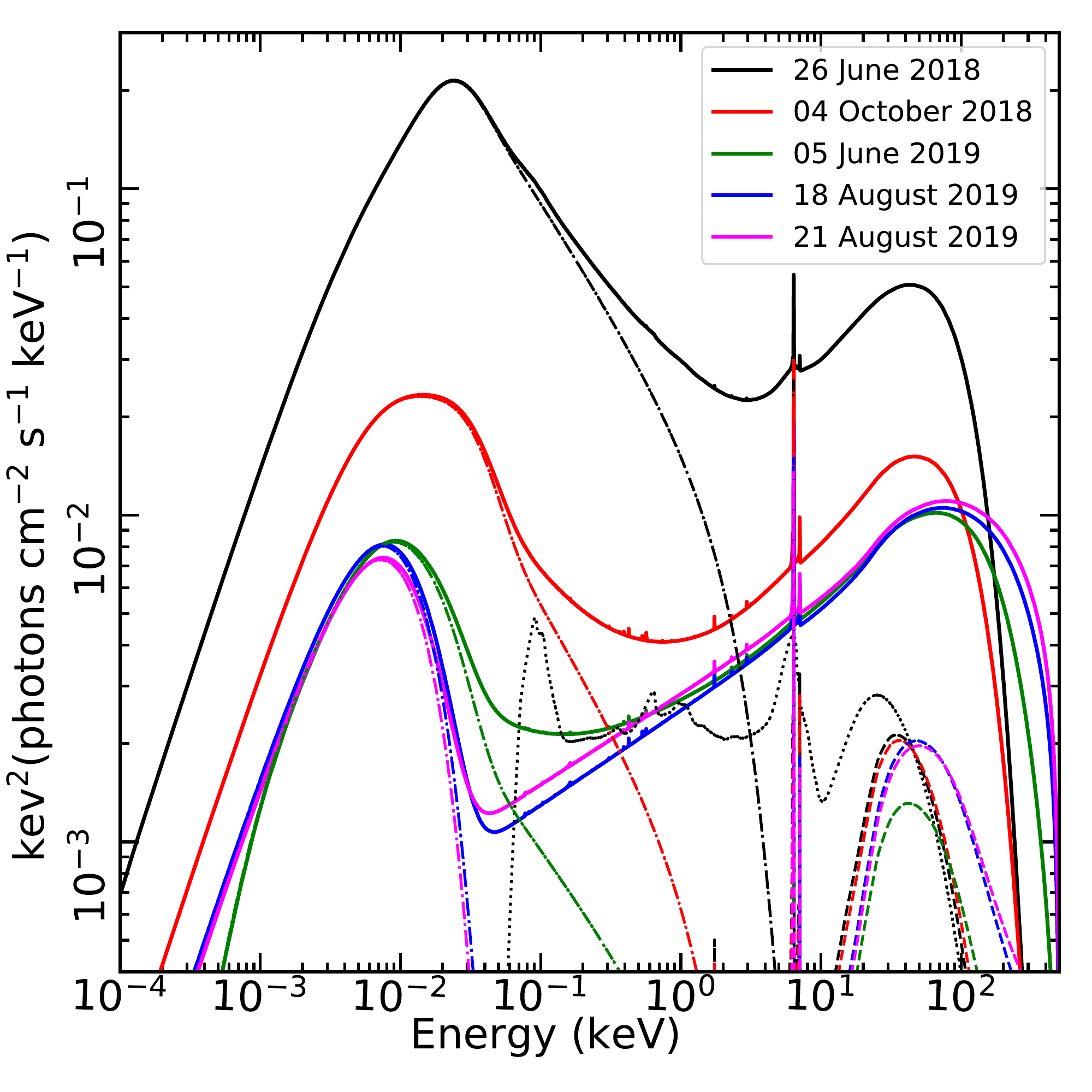}
    \caption{The best-fit models of NGC~1566 extrapolated in the $0.1\ev{}-500\kev{}$ band for 26 June 2018 (black), 04 October 2018 (red), 05 June 2019 (green), 18 August 2019 (blue), and 21 August 2019 (magenta) epochs derived from the best-fit data shown in Figure~\ref{fig_sed}. The total models (solid curves), accretion disk plus soft X-ray excess ({\textsc{optxagnf}} in dot-dashed curves), the blurred reflection component ({\textsc{relxill}} in  dotted curves), and the distant reflection components ({\textsc{xillver}} in  dashed curves) are also shown.}
    \label{fig_sed_model}
\end{figure}

In Fig.~\ref{fig_sed_model}, we show the best-fitting spectral models (after removing the absorption components)  derived from the five NUV/X-ray spectral data sets and extrapolated to $0.1\ev{}-500\kev{}$ bands. Clearly, the accretion disk, the soft X-ray excess, and the X-ray power-law components  rapidly decreased during the declining phase of the outburst (see Fig.~\ref{fig_sed_model}). In particular, the soft X-ray excess is more variable than the disk and the X-ray power-law components, which is consistent with our previous result \citep[see][]{Tripathi_&_Dewangan}. The soft X-ray excess was maximum in 26 June 2018, which decreased in 04 October 2018, and became negligible in 2019 epochs. { The decreasing soft X-ray excess with the decreasing accretion rate has been observed in other Seyferts \citep[see][]{2016A&A...588A..70B}}. The temperature of the warm corona responsible for the soft excess decreased from $kT_{w}\sim 0.57$ to $\sim 0.35\kev$, and the optical depth increased from $\tau_{w}\sim9$ to $\sim 12$. Also, the radius of the warm corona decreased during the declining phase of the outburst. These results suggest that at least part of the warm corona converted to the standard disk in the inner regions.  We have discussed the origin of the soft excess in terms of formation of the warm corona during the outburst in our previous work \citep[see][]{Tripathi_&_Dewangan}. The extreme multi-wavelength variability 
of NGC~1566 has been interpreted with the radiation pressure instability in the inner disk \citep[see][]{2019MNRAS.483L..88P, 2020A&A...641A.167S, Tripathi_&_Dewangan}.

\begin{figure}
    \centering
   \includegraphics[scale=0.42]{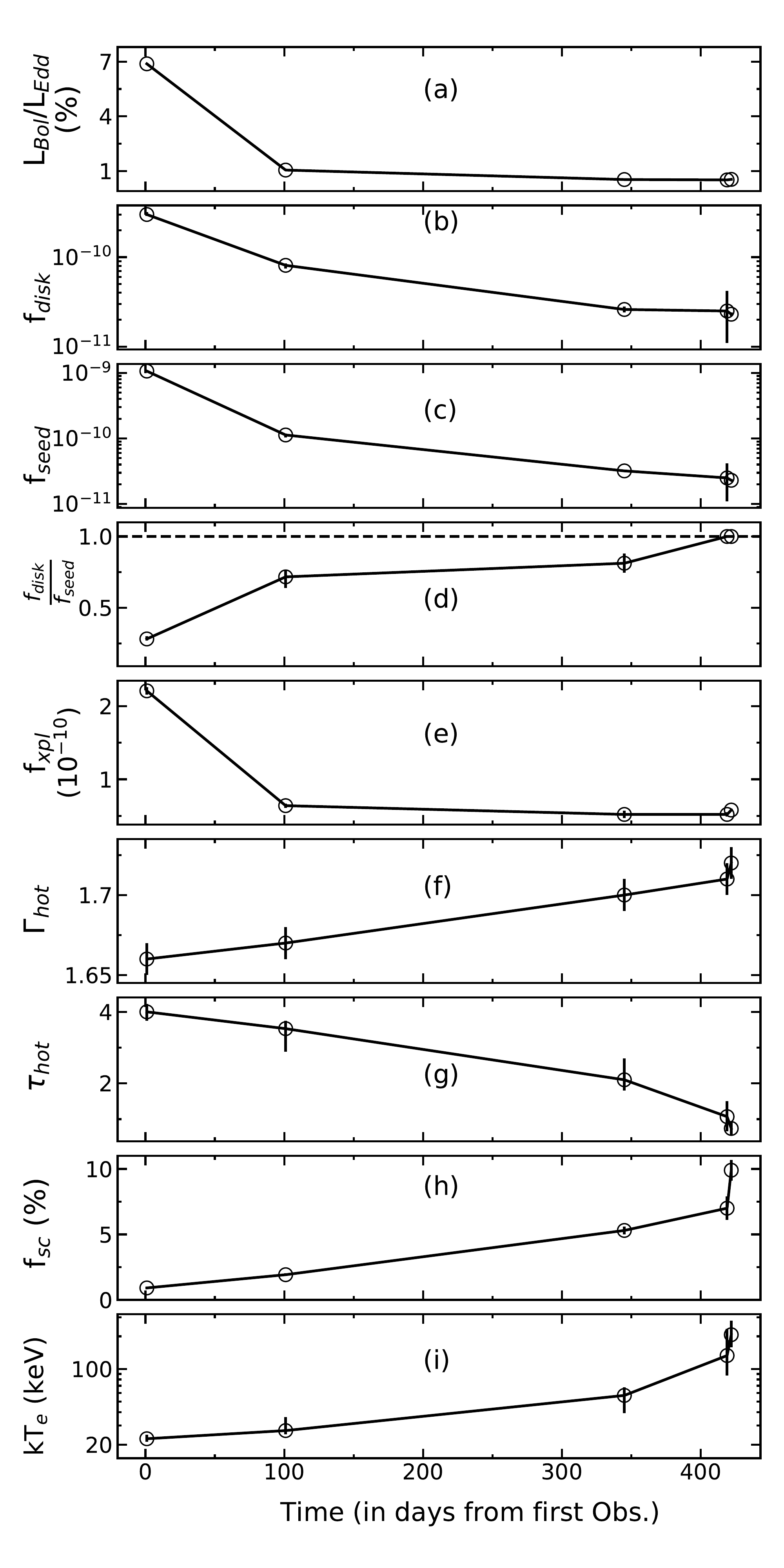}
    \caption{Variation in the spectral parameters and the fluxes derived from the spectral analyses of the five epochs datasets of NGC~1566 (see \S~\ref{sec_spectral} and Table~\ref{tab_nuvxrayspec})  (a) $L_{Bol}/L_{Edd} (\%)$, (b) $f_{disk}$, (c) $f_{seed}$, (d) ratio of the disk to total seed (disk + soft excess) flux $\frac{f_{disk}}{f_{seed}}$, (e) $f_{xpl}$ ($10^{-10}$), (f) $\Gamma_{hot}$, (g) $\tau_{hot}$, (h) $f_{sc} (\%)$, and (i) $kT_e$ (keV). The fluxes $f_{seed}$, $f_{xpl}$, and $f_{disk}$ are in the unit of \ergs{}\cm$^{-2}$\s$^{-1}$. The dotted line in the panel (d) represents $\frac{f_{disk}}{f_{seed}}=1$. For August 2019 epochs $f_{disk} = f_{seed}$, 
    therefore we have not shown the errors on the ratio}
    \label{fig_lcurve}
\end{figure}

\begin{figure*}
    \centering
    \includegraphics[scale=0.53]{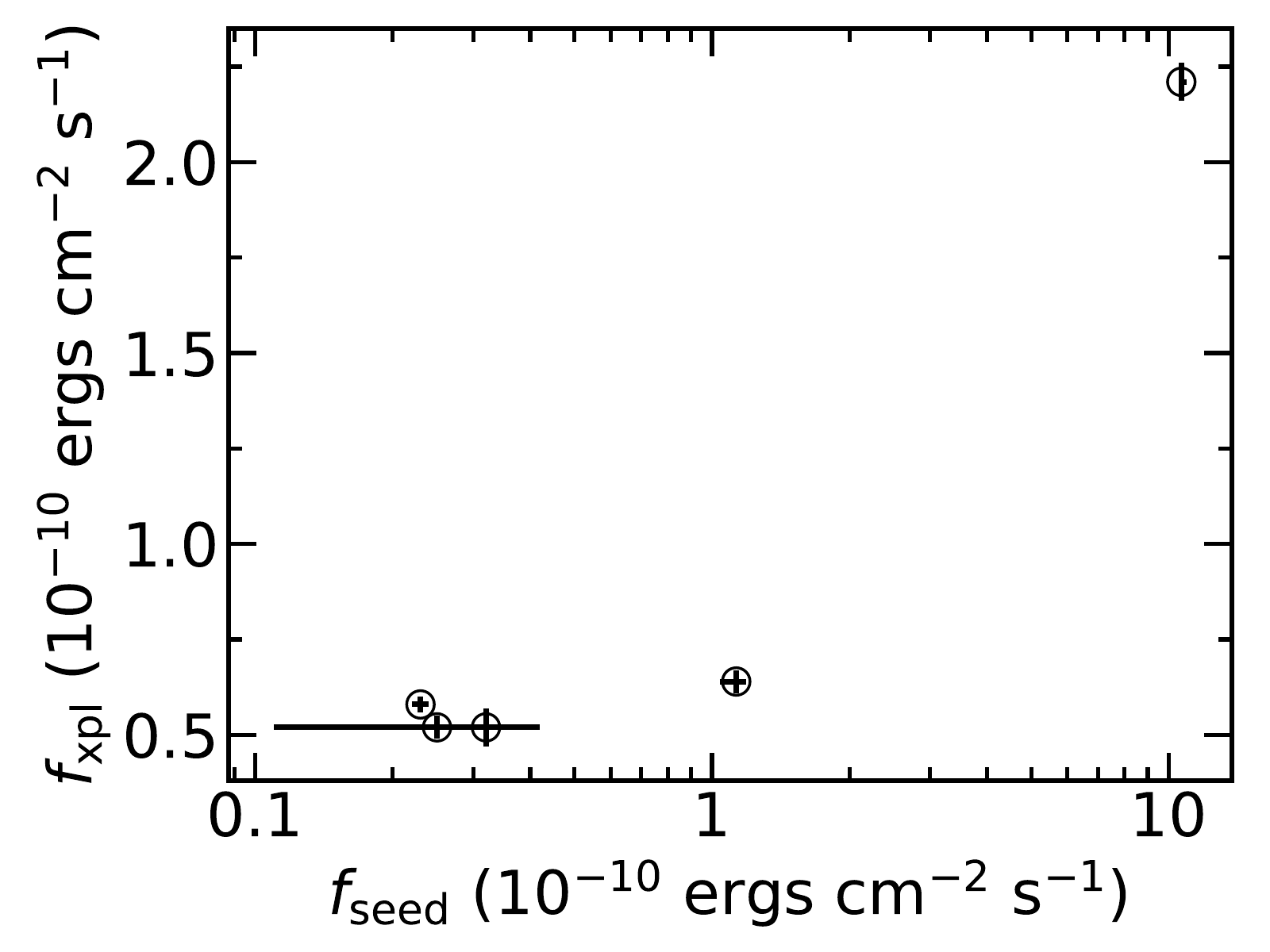}
   \includegraphics[scale=0.53]{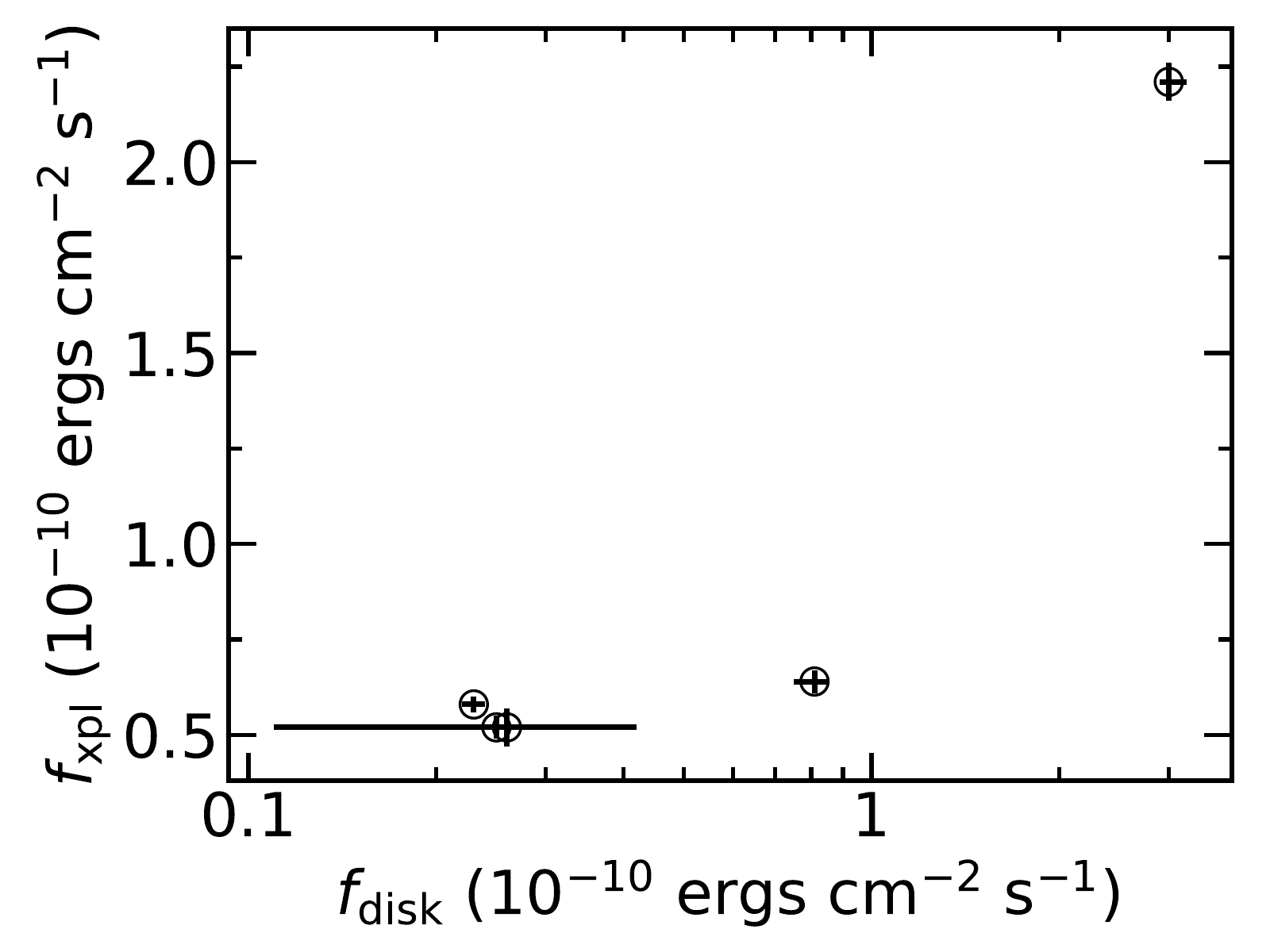}
   \includegraphics[scale=0.53]{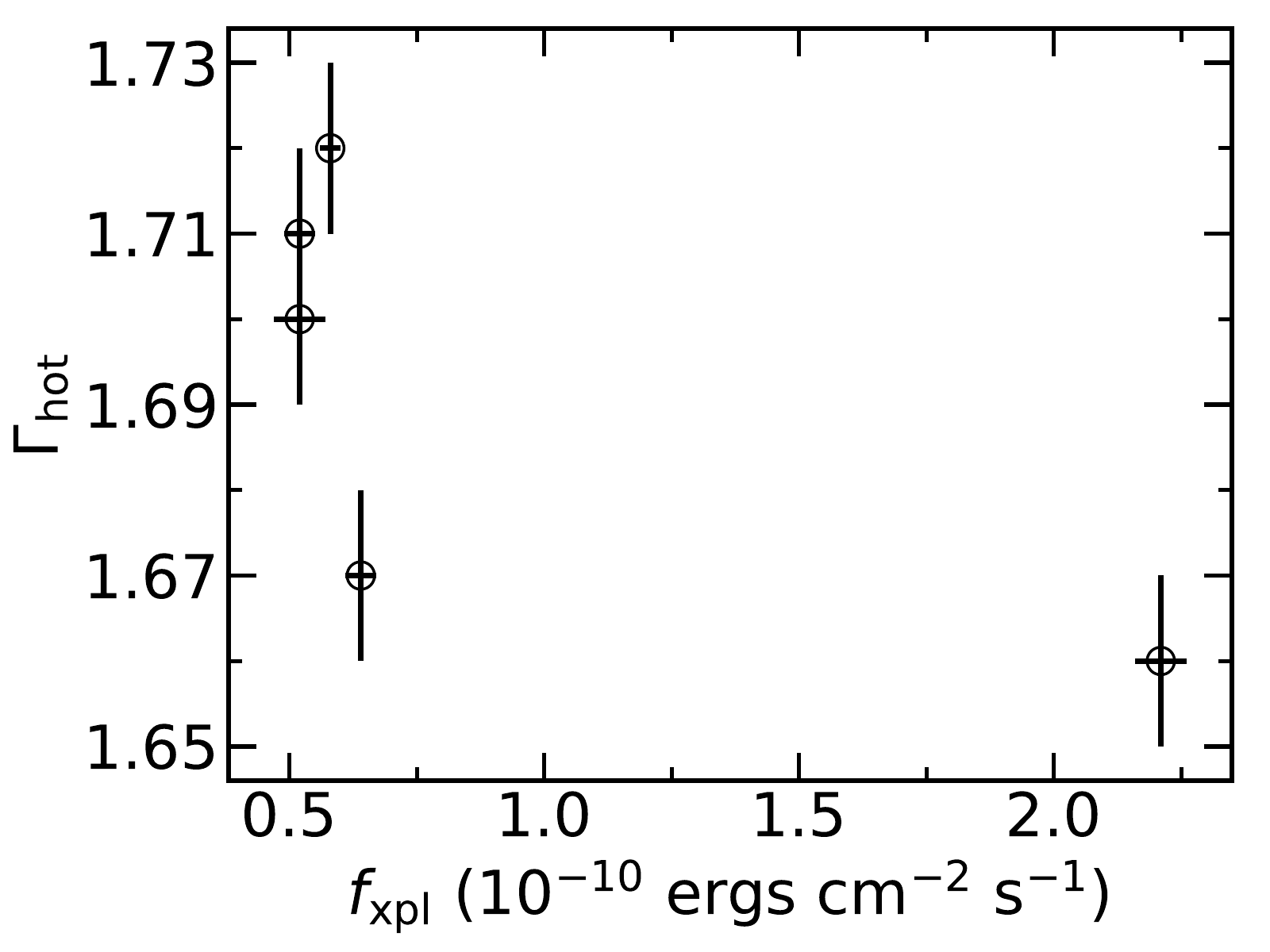}\\
   \includegraphics[scale=0.53]{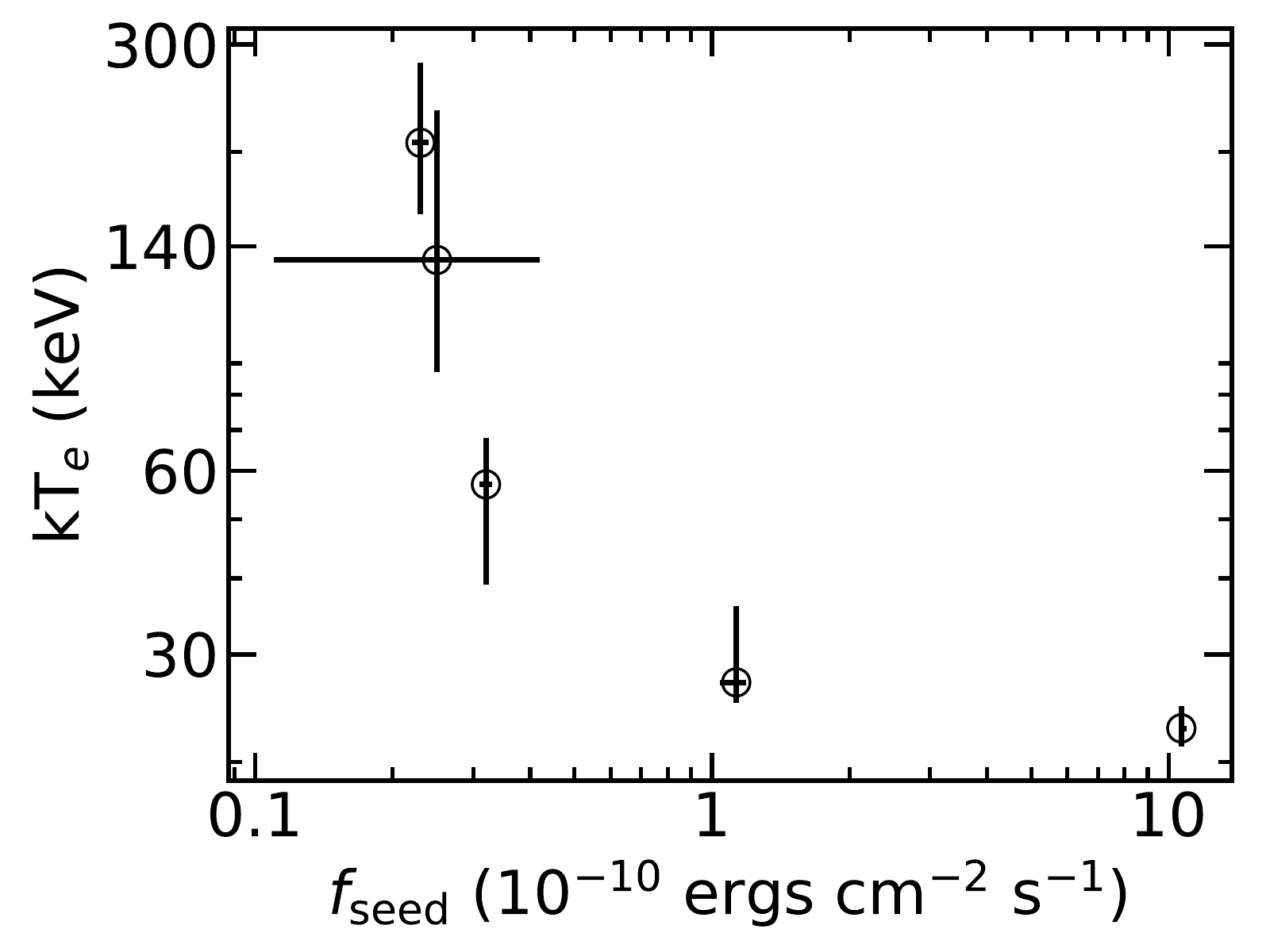}
   \includegraphics[scale=0.53]{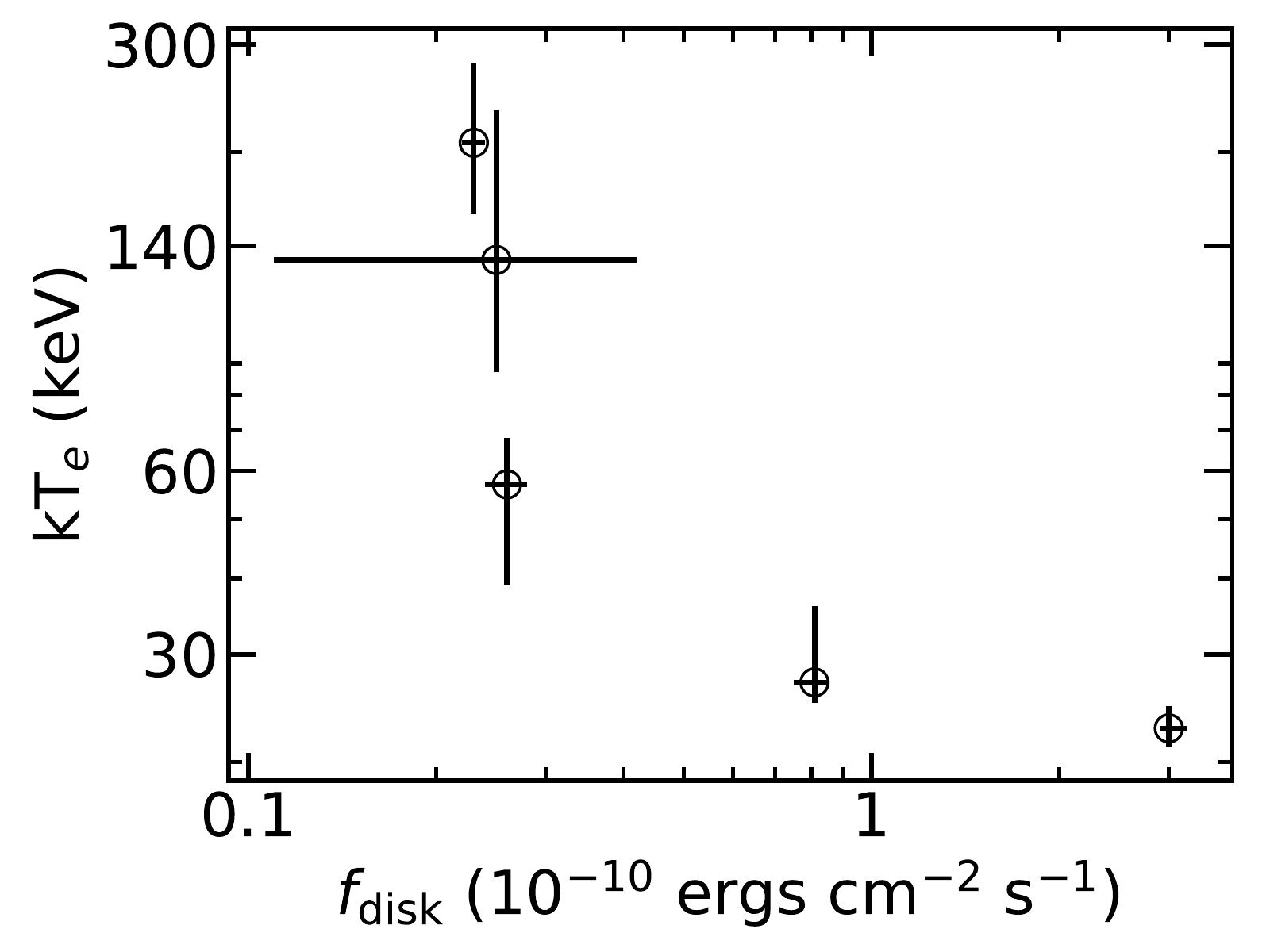}
   \caption{Top panels: The broadband (2--500\kev{}) X-ray power-law flux ($f_{xpl}$) with $f_{seed}$ (middle left) and  $f_{disk}$ (middle right). Middle panel: The photon-index ($\Gamma_{hot}$) with 
    $f_{xpl}$. Bottom panels: the electron temperature ($kT_e$) with $f_{seed}$ (bottom left) and $f_{disk}$ (bottom right)}

    \label{fig_var}
\end{figure*}

In Fig.~\ref{fig_lcurve}, we show variations in the best-fitting spectral parameters with time  relative to the first observation, 26 June 2018. Fig.~\ref{fig_lcurve}(a) shows the variations in the Eddington ratio $L_{Bol}/L_{Edd}$ that decreased from $\sim 7\%$ (at the outburst peak) to $\sim 0.5\%$ in $\sim 400$ days in the declining phase. The variable Eddington ratio during the outburst has been discussed in the context of the radiation pressure instability in the disk \citep[see][]{2019MNRAS.483L..88P, 2020A&A...641A.167S, Tripathi_&_Dewangan}. In the (b) and (c) panels of Fig.~\ref{fig_lcurve}, we have shown the seed (disk plus soft excess) flux ($f_{seed}$) and the standard disk flux ($f_{disk}$), respectively. The $f_{seed}$ and $f_{disk}$ decreased by factors of $\sim 43$ and $\sim 13$.  This implies that the variability in the $f_{seed}$ was mostly dominated by the soft X-ray excess component. 
This is much clear in Fig.~\ref{fig_lcurve}(d) where we have shown the ratio of the disk flux to the total seed flux ($f_{disk}/f_{seed}$). This ratio increased with time, and saturated at $f_{disk}/f_{seed} =1$ (shown as the black horizontal dashed line) during the August 2019 observations. This indicates that during the high flux states, 
the soft excess and the disk emission both provided the seed photon for thermal Comptonization in the hot corona, but a larger contribution came from the soft excess than the pure disk emission which made the corona to be sufficiently cooler at the outburst-peak ($\sim 22\kev{}$). The strong soft excess may be responsible for the low temperature of the hot corona in some narrow-line Seyfert 1 galaxies e.g., $kT_e\sim15 \kev$ in Akn~564 \citep{2017MNRAS.468.3489K,2020MNRAS.492.3041B}, $kT_e\sim 3-22\kev$ in IRAS~04416+1215 \citep{2022MNRAS.509.3599T}, $kT_e\sim 21-31\kev$ in Mrk~110 \citep{2021A&A...654A..89P}, $kT_e\sim 20\kev$ in ESO 362-G18 \citep{2021ApJ...913...13X}.
The broadband X-ray power-law flux in the 2--500\kev{} band ($f_{xpl}$)  decreased by a factor of $\sim 4$ from June 2018 to August 2019 (Fig.~\ref{fig_lcurve}(e). The variability of $f_{xpl}$ is  well correlated with the variability of $f_{seed}$ and $f_{disk}$ (see the panels (b), (c) and (e) of Fig.~\ref{fig_lcurve}). 
Fig.~\ref{fig_var}) shows relationship between different parameters. The $f_{xpl}$ showed exactly the same trends with $f_{seed}$ and $f_{disk}$. 
The decreasing X-ray power-law flux with the seed flux can be interpreted as due to thermal Comptonization of the seed photons in the hot corona where the decreasing number of seed photons is not able to cool  the corona effectively that  also results in decreasing  number of Comptonized X-ray photons \citep[see][]{2000ApJ...544..734N,Tripathi_2021}. The cooling or heating of the corona can also affects the shape of the Comptonized X-ray photons according to Equation~\ref{eqn1}. 
In case of NGC~1566, we found that the X-ray power-law photon-index was only weakly variable by $\Delta{\Gamma_{hot}}\leq0.06$ (see Fig.~\ref{fig_lcurve}(f)), and it was not noticeably correlated with the $f_{xpl}$ (see the middle panel in Fig.~\ref{fig_var}). 
On the other hand,
the optical depth of the corona $\tau_{hot}$ decreased from $\sim 4$ to $\sim 0.7$ (see Fig.~\ref{fig_lcurve}(g)), the covering fraction ($f_{sc}$) increased from $\sim 1\%$ to $\sim 10\%$ (see Fig.~\ref{fig_lcurve}(h)), and the electron temperature ($kT_e$) of the corona increased from $\sim 22\kev{}$ to $\sim 200\kev{}$ (see Fig.~\ref{fig_lcurve}(i)) during the declining phase of the outburst. Thus, the increasing electron temperature and decreasing optical depth of the hot corona resulted in only weakly varable the photon-index according to Eqn.~1.  Further, the bottom panels in Fig.~\ref{fig_var} show that the electron temperature of the corona was anti-correlated with the $f_{seed}$ and the $f_{disk}$ indicating that the corona became hotter with the decreasing seed photons. This is a clear evidence for thermal Comptonization  in the hot corona.

\begin{figure}
    \centering
     \includegraphics[scale=0.35]{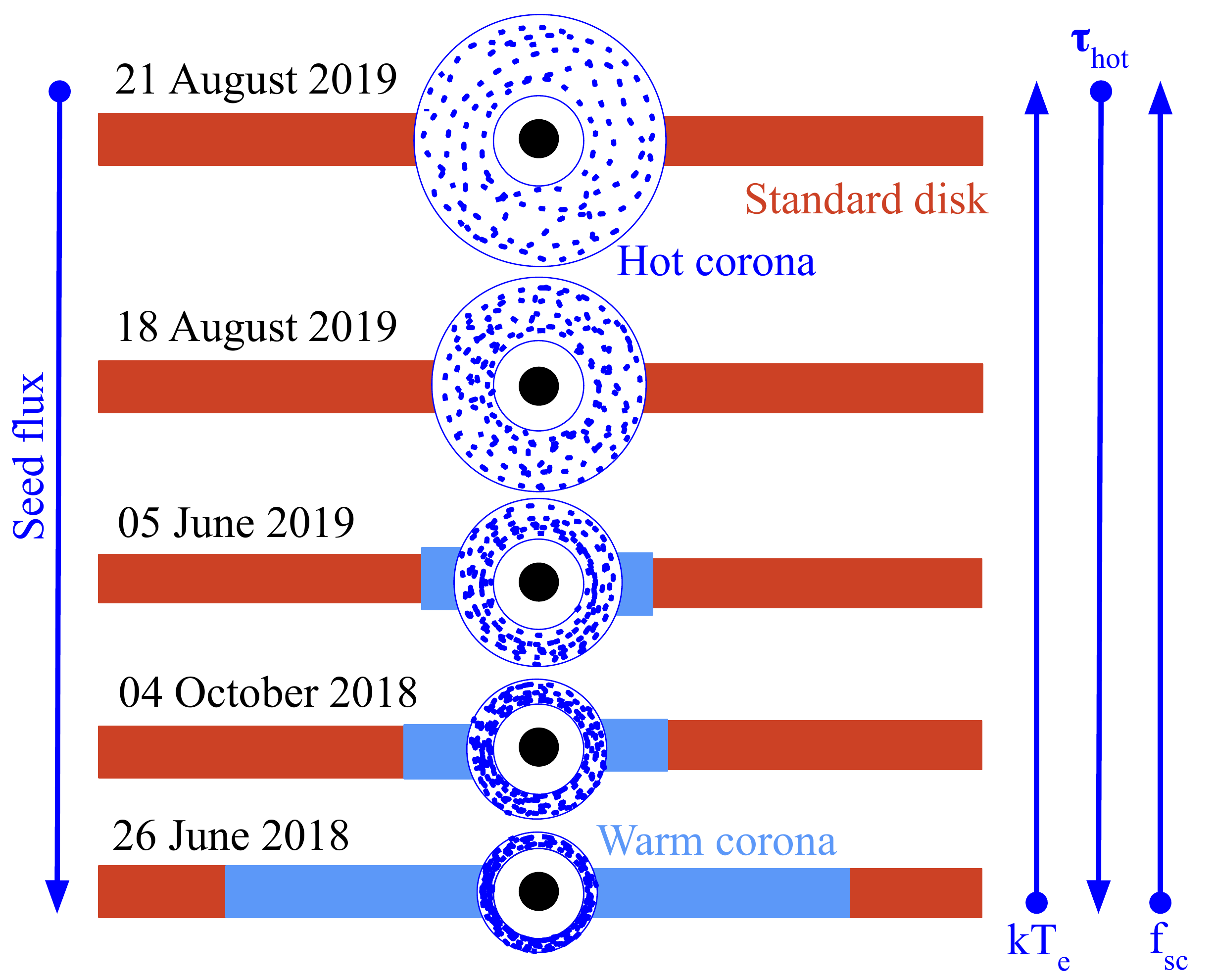}\\
   \caption{Schematic representation of possible evolution of the warm and hot corona in NGC~1566 during declining phase of the 2018 outburst. The dots inside the hot spherical corona represent the electrons. The variations in the coronal parameters are shown as  arrows pointing in the direction of increase. }
      \label{fig_corona}
\end{figure}

The observed variations in the scattering fraction $f_{sc}$ and the optical depth $\tau_{hot}$ suggest changes in the the geometry  of the hot corona during the declining phase of NGC~1566. The scattering fraction depends on the electron density and  the size of the hot corona. A bigger corona will intercept a larger fraction of the seed photons resulting in a larger scattering fraction $f_{sc}$. Similarly a larger electron density will provide a  larger number of targets for the inverse Compton scattering of seed photons which will result in larger $f_{sc}$. Since the decreasing $L_{Bol}/L_{Edd}$ resulted in the declining soft excess emission, presumably due to the shrinking warm corona, it is likely that the size of the hot corona increased at the expense of shrinking soft excess emitting region during the declining phase, hence resulting in increasing $f_{sc}$. However, with the increasing size, the optical depth of the hot corona  will also increase unless the electron density decreases.  Most likely  the decreasing accretion rate also resulted in decreasing fraction of accretion power fed to the corona, which was sufficient to maintain the electron density with increasing corona size. This ultimately may have caused in a faster decrease in the electron density than the increase in the corona size both due to decreasing accretion power and increasing size of the corona, and thus resulting in decreasing optical depth as observed.  In Fig.~\ref{fig_corona}, we show a schematic picture of possible evolution of the warm (light blue bar) and hot (central spherical region)) corona. The electrons randomly distributed in the hot corona are shown as  dark blue dots that decrease from June 2018 to 21 August 2019.
With decreasing accretion rate, the disk flux, the soft excess flux and the seed photons all decreased during the declining phase of NGC~1566. The decreasing number of seed photons then was not able to cool the hot corona, and most likely was responsible for the increasing temperature of the hot corona as observed (see Fig.~\ref{fig_corona} and the bottom panels of Fig.~\ref{fig_var}). The combination of a decreasing optical depth and increasing temperature of the corona then resulted in only minor change in the photon index of the X-ray power law (see Fig.~\ref{fig_lcurve}~[f] and equation~\ref{eqn1}). { In the high flux states of NGC~1566, the presence of soft X-ray excess suggests that the innermost regions of the disk is filled by the optically thick material in the form of a warm corona. 
The optically thick warm corona can be considered as the inner part of the accretion disk. In that case, in the high flux state, the accretion disk seems to extend down to the innermost regions. This state is similar to the high/soft state of black hole X-ray binaries (BHBs) in which thermal emission from the disk is dominated \citep[see][]{2016ASSL..440...61B}. Whereas,} in the lowest flux states of NGC~1566 in August 2019, the standard disk without the soft excess emitting region appears to be truncated at $\sim 20r_g$, and the corona is most likely filling the innermost regions below the truncation radius, and the source may be evolving with decreasing accretion rate towards a state similar to the  low/hard state of BHBs. The emission from the low/hard state of BHBs is dominated by  X-ray power law components with reduced or negligible thermal emission from the accretion disk. It is believed that the accretion disks in these objects is most likely truncated at large radii \citep[see][]{2001ApJ...555..477M, 2001ApJ...555..483E, 2016ASSL..440...61B}. 


Another possibility to explain the increasing $f_{sc}$ and decreasing $\tau$ is to decrease the size of the hot corona and increase the electron density faster than the decrease in the size of the hot corona which may require increase in accretion power fed to the corona. This is unlikely as the accretion power is observed to be decreasing during the declining phase.

\section{conclusion}
\label{sec_conclude}

The main results of our broadband UV to X-ray spectral variability study of NGC~1566 are as follows.
\begin{itemize}
\item The accretion disk, soft X-ray excess, and the X-ray power-law components were extremely variable during the declining period of the outburst. The soft excess was maximum in June 2018, which reduced in October 2018, and became negligible in 2019 epochs. 
\item The broadband power-law flux was correlated with the soft excess plus disk flux suggesting  thermal Comptonization of both  the soft excess and the disk photons in the hot corona.
 \item At the high flux levels when the soft excess was present, 
both the soft excess and the disk components provided the seed photons for thermal Comptonization in the hot corona, whereas at the low flux levels when the soft excess was absent, the pure disk emission alone provided the seed photons. 
\item We found that the photon-index of the X-ray power-law was only weakly variable ($\Delta{\Gamma}\leq 0.06$), and it was not well correlated with the X-ray power-law flux.
\item The electron temperature of the corona increased from $\sim 22\kev{}$ to $\sim 200\kev{}$ with decreasing seed flux from June 2018 to August 2019. The low temperature of the corona at the outburst peak is most likely due to the cooling effect of the strong soft excess, and the disappearance of the soft excess and the decreasing disk flux most likely could not cool the corona effectively.

\item The scattering fraction increased from $\sim 1\%$ to $\sim 10\%$ during the declining phase of the outburst. We suggest that increasing size of the corona is mainly responsible for it.

\item The optical depth of the hot corona decreased from $\sim 4$ to $\sim 0.7$. Most likely the decreasing number of electrons in the corona  due to the decreasing accretion power is responsible for the observed variation in the optical depth.

\end{itemize}

\acknowledgments

This research has made use of archival data of \xmm{} and the SAS software provided by the \xmm{} Science Archive developed by the ESAC Science Data Centre (ESDC) with requirements provided by the \xmm{} Science Operations Centre. This research has made use of archival data of \swift{} and \nustar{} observatories via High Energy Astrophysics Science Archive Research Center Online Service, provided by the NASA Goddard Space Flight Center. This research has made use of {\it HEASoft} software. This research has made use of the {\it Python} packages. This research has made use of the NASA/IPAC Extragalactic Database (NED), which is operated by the Jet Propulsion Laboratory, California Institute of Technology, under contract with the National Aeronautics and Space Administration (NASA). 
 
\facilities{\xmm{}, \nustar{}, \swift{}}

\software{
    HEAsoft \citep{2014ascl.soft08004N}
	XSPEC \citep{1996ASPC..101...17A},
	SAOImageDS9 \citep{2003ASPC..295..489J},
		SAS (v18.0.0; \citealt{2004ASPC..314..759G}) 
}


\bibliography{mybib}
\bibliographystyle{aasjournal}

\appendix
\counterwithin{figure}{section}
\section{Additional figures}
\begin{figure}
    \centering
     \includegraphics[scale=0.55]{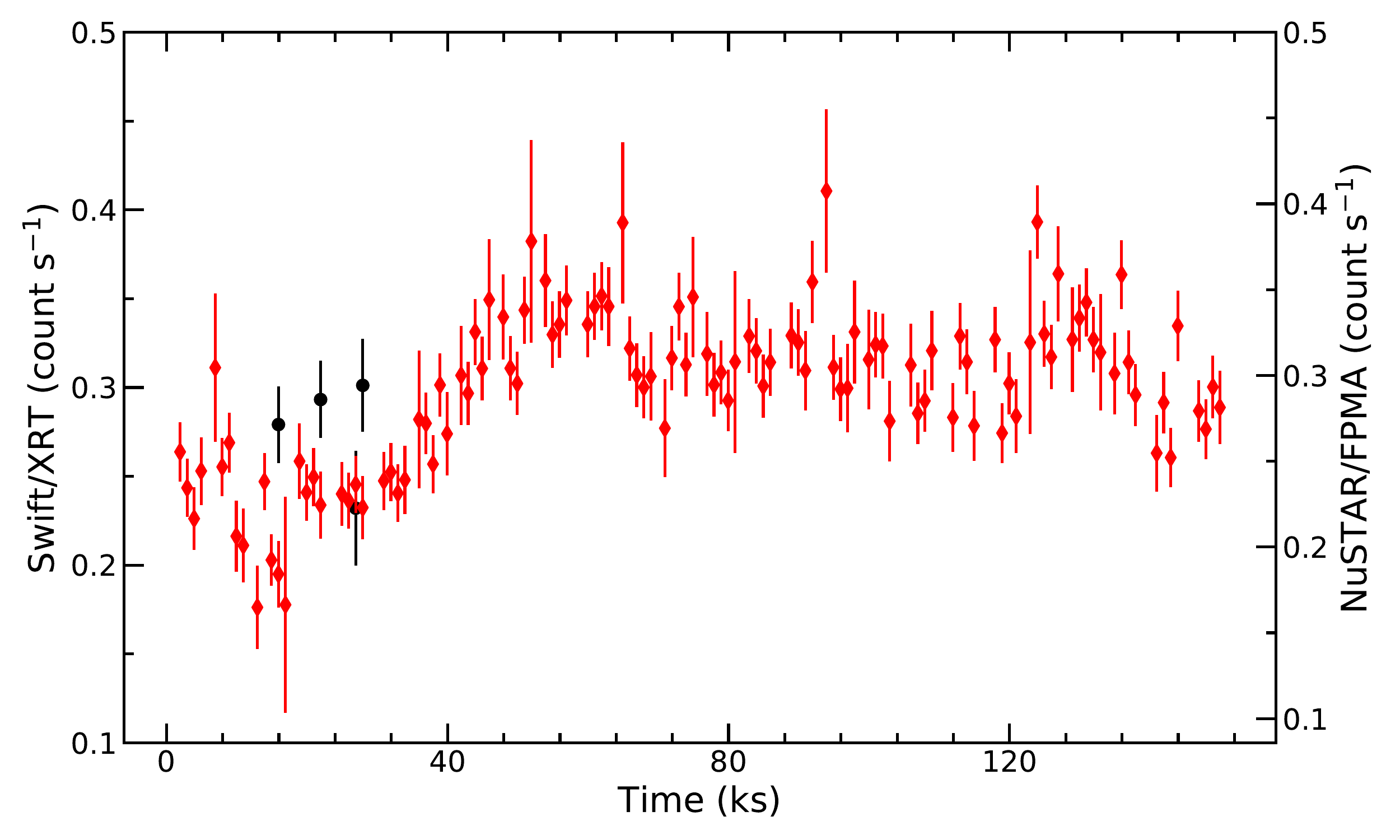}
    \caption{\swift{}/XRT (black filled circles) and \nustar{}/FPMA (red diamonds) lightcurves for 21 August 2019} 
    \label{fig_lc}
\end{figure}

\begin{figure}
    \centering
    \includegraphics[scale=0.36]{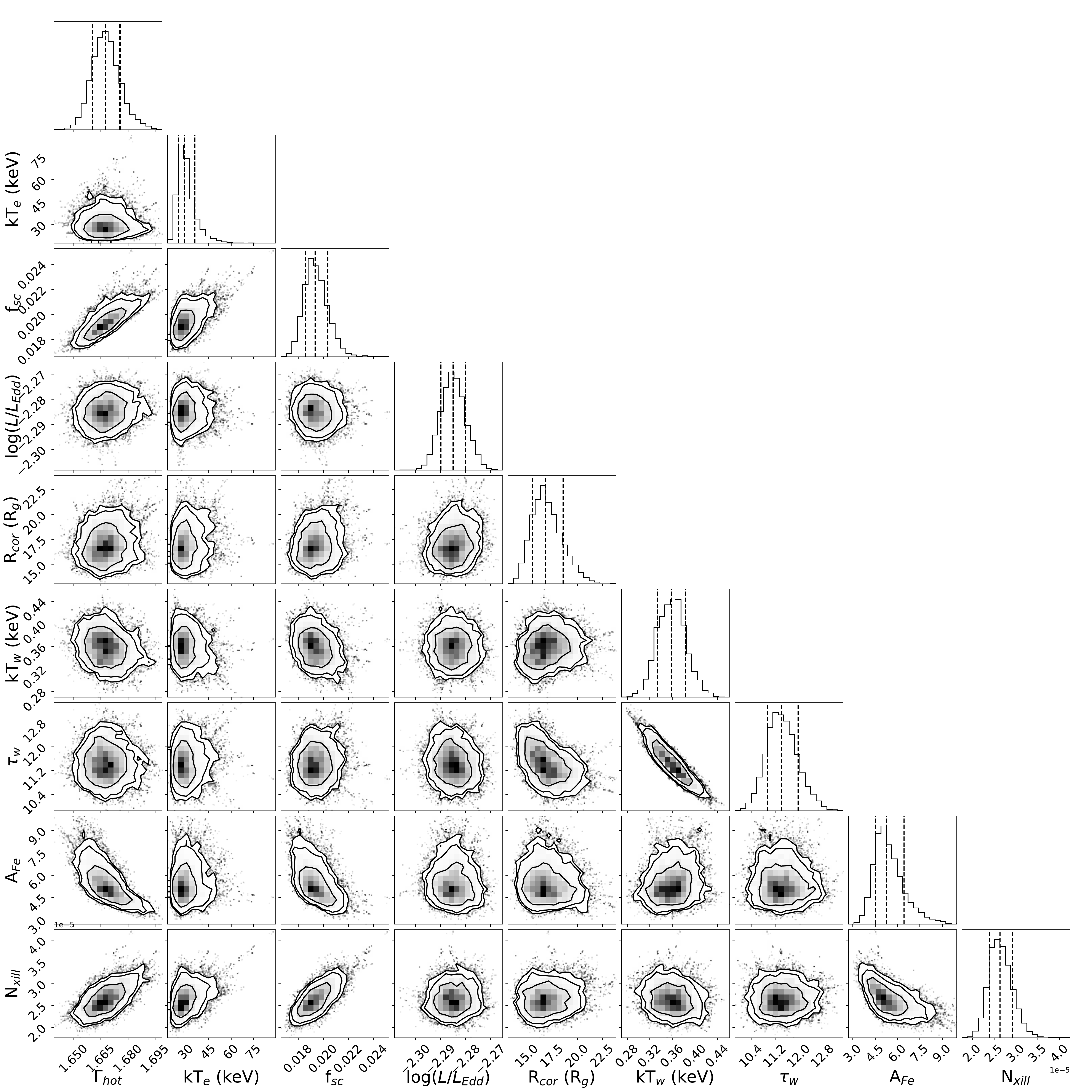}
    \caption{Contour plots in 68\%, 90\%, and 95\% confidence levels derived from the MCMC chain for 04 October 2018 epoch.}
    \label{fig_mcmc_contours}
\end{figure}

\end{document}